\documentclass[aps,pra,twocolumn,showpacs]{revtex4-2}
\usepackage{epsfig,amssymb,amsfonts}
\usepackage{amsmath,amsfonts,amssymb}
\usepackage{graphicx}
\usepackage{caption}
\usepackage{xcolor}
\input{epsf}
\usepackage{epstopdf}
\usepackage{natbib}
\usepackage{braket}

\usepackage[labelformat=simple]{subcaption}
\makeatletter
\renewcommand\p@subfigure{\thefigure}
\makeatother

\begin{document}

\title{Fast charging of an Ising spin pair quantum battery using optimal control}
\author{Vasileios Evangelakos}
\author{Emmanuel Paspalakis}
\author{Dionisis Stefanatos}
\email{dionisis@post.harvard.edu}
\affiliation{Materials Science Department, School of Natural Sciences, University of Patras, Patras 26504, Greece}
\date{\today}
\begin{abstract}

We consider the problem of maximizing the stored energy for a given charging duration in a quantum battery composed of a pair of spins-$1/2$ with Ising coupling starting from the spin-down state, using bounded transverse field control. We map this problem to an optimal control problem on a single qubit and using optimal control theory we show that, although a single bang pulse can quickly achieve considerable charging levels for relatively large upper control bounds, higher levels of stored energy including complete charging are accomplished by a bang-singular-bang pulse-sequence, where the intermediate singular pulse is an Off pulse. If the control is restricted between zero and a maximum value, the initial and final bang pulses attain the maximum bound but have different durations, while if it is restricted between symmetric negative and positive boundaries, the bang pulses have the same duration but opposite boundary values. For both cases we provide transcendental equations from which the durations of the individual pulses in the optimal pulse-sequence can be calculated. For the case of full charging we surprisingly find that the three ``switching" functions for the equivalent qubit problem become zero while the adjoint ket does not, in consistency with optimal control theory.% The corresponding optimal solution is abnormal and contains a singular arc. This latter behavior is attributed to that the formulation of the problem on the qubit system imposes a requirement on the global phase and not just on the position of the Bloch vector on the sphere, as is usually the case. 

\end{abstract}

\maketitle

\section{Introduction}

Quantum batteries (QBs) \cite{AlickiFannes} constitute a rapidly growing research field within modern quantum science and technology \cite{campaioli2023colloquium}. Theoretical studies suggest that they can exploit the peculiar features of quantum mechanics, like superposition and entanglement, for energy storage and retrieval at a microscopic scale, with superior performance relative to their classical counterparts \cite{Binder_2015,PhysRevLett.125.236402,PhysRevLett.128.140501,PhysRevLett.118.150601,10.1116/5.0184903}. In spite of the many challenges \cite{PhysRevResearch.2.023113}, various proposals for QBs are currently under investigation, for example two- and three-level systems, many-body systems like spin chains, the Su–Schrieffer–Heeger and Sachdev–Ye–Kitaev models, as well as quantum cavities and oscillators \cite{PhysRevB.98.205423,PhysRevA.97.022106,PhysRevE.99.052106,PhysRevLett.122.210601,PhysRevE.100.032107,PhysRevLett.122.047702,PhysRevB.102.245407,PhysRevE.101.062114,PhysRevE.103.042118,PhysRevE.105.064119,PhysRevB.105.115405,Barra_2022,Shaghaghi_2022,PhysRevA.108.042618,a19b2d5b090c43b4aef8ac7dca563502,Downing2023,hadipour2023practical,PhysRevE.102.052109,PhysRevA.100.043833,10.3389/fphy.2022.1097564,Crescente2020,PhysRevA.106.042601,PhysRevA.104.042209,Niedenzu2018,PhysRevLett.120.117702,PhysRevA.103.033715,PhysRevB.100.115142,PhysRevE.104.054117,PhysRevResearch.4.013172,Hu_2022,wenniger2023experimental,doi:10.1126/sciadv.abk3160,Dou_2020,Moraes_2021,Dou2021,dou2022charging,PhysRevA.109.032201,PhysRevA.107.032218,Rodríguez_2024}. Regarding experimental implementation, some prototype QB models have been recently realized in various platforms, for example superconductors \cite{Hu_2022}, quantum dots \cite{wenniger2023experimental}, organic microcavities \cite{doi:10.1126/sciadv.abk3160}, and systems of nuclear spins \cite{PhysRevA.106.042601}.

A crucial task in a QB system is the charging process used to increase the stored energy in the battery. Various quantum control methods \cite{Stefanatos_2020} have been proposed for the efficient implementation of the charging procedure. Stimulated Raman adiabatic passage (STIRAP) \cite{RevModPhys.79.53,RevModPhys.89.015006} has been employed for charging a three-level quantum battery, in open \cite{PhysRevE.100.032107} and closed \cite{Dou_2020} configurations, while adiabatic passage has been used for a QB composed of two noninteracting qubits \cite{PhysRevE.101.062114}. Shortcuts to adiabaticity \cite{RevModPhys.91.045001} have been exploited to accelerate charging for both systems \cite{Dou2021,Moraes_2021}, as well as for a Lipkin-Meshkov-Glick QB \cite{dou2022charging}, while the closely related quantum adiabatic brachistochrone approach has been used in the recent experiment involving a superconducting qutrit QB \cite{Hu_2022}. A strong pulse was employed for the fast charging of an interacting many-body QB \cite{PhysRevA.109.032201}, and the quantum speed limit formalism \cite{Deffner_2017} for finding tighter bounds on the charging time \cite{shrimali2024stronger}. Recently, optimal control theory has been utilized to find the optimal drive for the fast charging of a qubit QB \cite{PhysRevA.107.032218}, while numerical optimal control has been applied for the efficient manipulation of a QB composed of two coupled oscillators \cite{Rodríguez_2024}.

Following the spirit of Ref. \cite{PhysRevA.107.032218}, in this work we consider a quantum battery consisting of a pair of spins-$1/2$ with Ising coupling, in the usual nuclear magnetic resonance framework with constant longitudinal field and bounded time-dependent transverse fields which act as the controls, and study the problem of maximizing the stored energy in the battery for a given charging duration and when starting from the spin-down state. We show that this problem can be mapped to an optimal control problem on an effective two-level system and use optimal control theory \cite{Boscain21,SugnyReview,Boscain06,Mason08,PhysRevA.108.062425,Stojanovic23,Khaneja01} to investigate it. It turns out that for short available durations a single bang pulse can quickly achieve considerable charging levels for relatively large upper control bounds. But when the charging duration exceeds a certain threshold, then we provide strong evidence (without a strict mathematical proof) that the optimal pulse-sequence takes the bang-singular-bang form, where the intermediate singular pulse is an Off pulse, and this sequence achieves complete charging of the battery (the spin-up state) in minimum time. If the control is restricted between zero and a maximum bound, the initial and final bang pulses attain the maximum value but have different durations, while if it is restricted between symmetric negative and positive boundaries, the bang pulses have the same duration but take opposite boundary values. The threshold duration beyond which the bang-singular-bang optimal solution arises is shorter for the latter case where the control can take also negative values. For both cases we provide transcendental equations from which the durations of the individual pulses in the bang-singular-bang pulse-sequence can be calculated, while note that the same stored energy is achieved if the initial and final bang pulses are interchanged. From a control theoretic point of view regarding the optimal control problem on the effective two-level system, we find that in the case of full charging the triplet of ``switching" functions surprisingly vanishes while the adjoint ket is nonzero, in consistency with optimal control theory. %This solution actually corresponds to abnormal optimal control and contains a singular arc. We attribute this latter phenomenon to the fact that the equivalent formulation of the problem places a requirement on the global phase of the qubit and not just on the position of the Bloch vector on the sphere, as is usually the case.   

The paper has the following structure. In Sec. \ref{sec:II} we formulate the optimal control problem of charging the considered spin-pair QB, in Sec. \ref{sec:Optimal_Solution} we use optimal control theory to derive optimal charging protocols, and in Sec. \ref{sec:bsb} we examine in detail the bang-singular-bang protocol. In Sec. \ref{sec:full} we present the optimal solution for full charging and provide specific examples in Sec. \ref{sec:results}, while Sec. \ref{sec:conclusion} concludes the present work.

\section{Charging a spin-pair quantum battery as an optimal control problem}
\label{sec:II}

We consider a quantum battery consisting of two spins-$1/2$ coupled with an Ising interaction along the $z$-direction. This Ising spin pair is embedded in the standard NMR framework, where the magnetic field acting on the spins has a constant longitudinal component and time-dependent transverse components which serve as the available controls. The corresponding Hamiltonian of the system is ($\hbar=1$)
\begin{equation}
\hat{H}(t) = \hat{H}_0+\hat{H}_1(t)
\end{equation}
with
\begin{subequations}
\begin{eqnarray}
    \hat{H}_0 & = & \frac{\Omega_z}{4}(\sigma_{1z}+\sigma_{2z}) + \frac{J}{2}\sigma_{1z}\sigma_{2z},  \\
    \hat{H}_1(t)       & = & \frac{\Omega_x(t)}{4} (\sigma_{1x}+\sigma_{2x}) + \frac{\Omega_y(t)}{4} (\sigma_{1y}+\sigma_{2y}),
\end{eqnarray}
\end{subequations}
where $J$ expresses the Ising interaction, $\Omega_z$ the constant longitudinal field and $\mathbf{\Omega}(t)=[\Omega_x(t), \Omega_y(t)]^T$ the transverse control field. 

It turns out \cite{Vitanov01} that the above Hamiltonian couples the triplet states
\begin{subequations}
\label{basis}
    \begin{eqnarray}
    \ket{\psi_0} &=& \ket{00}, \\
    \ket{\psi_1} &=& \frac{1}{\sqrt{2}} \left(\ket{01} + \ket{10}\right), \\
    \ket{\psi_2} &=& \ket{11},
    \end{eqnarray}
\end{subequations}
while the singlet state 
\begin{equation}
\label{singlet}
\ket{\varphi} = \frac{1}{\sqrt{2}} \left(\ket{01} - \ket{10}\right)
\end{equation}
is decoupled, where $\ket{0}=(0 \; 1)^T, \ket{1}=(1 \; 0)^T$ are the individual spin-down and spin-up states, respectively.
%\begin{subequations}
%\label{up_down}
%    \begin{eqnarray}
%    \ket{0} &=& \left(\begin{array}{c} 0 \\ 1 \end{array}\right) \\
%    \ket{1} &=& \left(\begin{array}{c} 1 \\ 0 \end{array}\right)
%    \end{eqnarray}
%\end{subequations}
If we express the state of the system when restricted on the triplet manifold as
\begin{equation}
    \ket{\psi(t)} = a_0(t) \ket{\psi_0} + a_1(t) \ket{\psi_1} + a_2(t) \ket{\psi_2}
\end{equation}
where $a_i, i=0, 1, 2$ are the amplitudes of the corresponding states, then from Schr\"odinger equation
\begin{equation}
\label{Schrodinger}
    i \frac{\partial}{\partial t} \ket{\psi(t)} = \hat H \ket{\psi(t)}
\end{equation}
we get the following equation for $\mathbf{a}=(a_0 \; a_1 \; a_2)^T$,
\begin{equation}
i\frac{d\mathbf{a}}{dt}
= \left[
\begin{array}{ccc}
 \frac{J -\Omega_z }{2} & \frac{\Omega_x(t)-i \Omega_y(t)}{2 \sqrt{2}} & 0 \\
 \frac{\Omega_x(t)+i \Omega_y(t)}{2 \sqrt{2}} & -\frac{J}{2} & \frac{\Omega_x(t)-i \Omega_y(t)}{2 \sqrt{2}} \\
 0 & \frac{\Omega_x(t)+i \Omega_y(t)}{2 \sqrt{2}} & \frac{J+\Omega_z}{2} \\
\end{array}
\right] 
\mathbf{a}.
\end{equation}

Charging of this quantum battery corresponds to transferring population from the initial spin-down state $\ket{\psi(0)}=\ket{\psi_0}$ to the excited states such that at the fixed final time $T$ the stored energy is maximized
\begin{subequations}
\begin{eqnarray}
    \Delta E &=& E(T) - E(0) \nonumber \\
             &=& \bra{\psi(T)}\hat{H}_0\ket{\psi(T)} - \bra{\psi(0)}\hat{H}_0\ket{\psi(0)}.
\end{eqnarray}
\end{subequations}
This definition leads to the following expression of the stored energy in terms of the final populations of the spin-up and spin-down states 
\begin{equation}
\label{E_stored}
\frac{\Delta E}{\Omega_z} = \left(\frac{1}{2}- \chi \right)\left(1-\left|a_0(T)\right|^2\right) + \left(\frac{1}{2}+ \chi \right)\left|a_2(T)\right|^2,
\end{equation}
where 
\begin{equation}
\label{ratio}
\chi = \frac{J}{\Omega_z} < \frac{1}{2}
\end{equation}
and we have used that the system is closed so $\left|a_1(T)\right|^2 = 1 - \left|a_0(T)\right|^2 - \left|a_2(T)\right|^2$. The inequality in Eq. (\ref{ratio}) assures that $\ket{\psi_0}$ is indeed the ground state of the system.

If we use the following transverse control field
\begin{subequations}
\label{set_omega}
    \begin{eqnarray}
    \Omega_x(t) &=& \Omega (t) \cos (\omega_c t),\\
    \Omega_y(t) &=& -\Omega (t) \sin (\omega_c t),
    \end{eqnarray}
\end{subequations}
where $\Omega(t)$ denotes the total field with central frequency $\omega_c$,
and make the transformation
\begin{subequations}
\label{alpha_to_c}
    \begin{eqnarray}
    c_0 &=& a_0 e^{i\left(\frac{J}{2}-\omega_c\right) t}, \\
    c_1 &=& a_1 e^{i\frac{J}{2} t}, \\
    c_2 &=& a_2 e^{i\left(\frac{J}{2}+\omega_c\right) t},
    \end{eqnarray}
\end{subequations}
then we get the following equation for $\mathbf{c}=(c_0 \; c_1 \; c_2)^T$,
\begin{equation}
\label{c_eq}
i \frac{d\mathbf{c}}{dt} = \left[
\begin{array}{ccc}
 \omega_c-\frac{\Omega_z}{2} & \frac{\Omega (t)}{2 \sqrt{2}} & 0 \\
 \frac{\Omega (t)}{2 \sqrt{2}} & -J  & \frac{\Omega (t)}{2 \sqrt{2}} \\
 0 & \frac{\Omega (t)}{2 \sqrt{2}} & \frac{\Omega_z}{2}-\omega_c \\
\end{array}
\right] \mathbf{c}.
\end{equation}
Note that this equation also describes the biexciton system in a quantum dot \cite{10.1063/5.0053859}, thus the present analysis can be also applied there.
Since we are interested in maximizing the stored energy and the spin-up state $\ket{\psi_2}$ is the one with highest energy, we set $\omega_c=\Omega_z/2$ and end up with the equation
% \begin{equation}
% \label{Schrodinger}
% H=\left(
% \begin{array}{ccc}
%  0 & \frac{\Omega (t)}{2 \sqrt{2}} & 0 \\
%  \frac{\Omega (t)}{2 \sqrt{2}} & -J -\frac{\Delta}{2}+\omega & \frac{\Omega (t)}{2 \sqrt{2}} \\
%  0 & \frac{\Omega (t)}{2 \sqrt{2}} & 2 \omega -\Delta \\
% \end{array}
% \right)
% \end{equation}
\begin{equation}
\label{Three_level}
i \frac{d\mathbf{c}}{dt} = \left[
\begin{array}{ccc}
 0 & \frac{\Omega (t)}{2 \sqrt{2}} & 0 \\
 \frac{\Omega (t)}{2 \sqrt{2}} & -J  & \frac{\Omega (t)}{2 \sqrt{2}} \\
 0 & \frac{\Omega (t)}{2 \sqrt{2}} & 0 \\
\end{array}
\right] \mathbf{c}.
\end{equation}

% \begin{equation}
% \label{H}
% H
% =
% \left(
% \begin{array}{ccc}
%  0 & \frac{\Omega (t)}{2 \sqrt{2}} & 0 \\
%  \frac{\Omega (t)}{2 \sqrt{2}} & -J  & \frac{\Omega (t)}{2 \sqrt{2}} \\
%  0 & \frac{\Omega (t)}{2 \sqrt{2}} & 0 \\
% \end{array}
% \right)
% \end{equation}

The dynamics described by Eq. (\ref{Three_level}) can be actually mapped to that of a spin-1/2 using the transformation
\begin{subequations}
\label{A_B_Gamma}
    \begin{eqnarray}
    A &=& \frac{c_2+c_0}{\sqrt{2}}, \\
    B &=& c_1, \\
    C &=& \frac{c_2-c_0}{\sqrt{2}}.
    \end{eqnarray}
\end{subequations}
We easily find that $\dot{C}=0$, while $A$,$B$ satisfy the two-level system equation
\begin{equation}
\label{two_level}
i
\left(
\begin{array}{c}
 \dot{A} \\
 \dot{B}
\end{array}
\right)
=
\left[
\begin{array}{cc}
 0 & \frac{\Omega (t)}{2} \\
 \frac{\Omega (t)}{2} & -J  \\
\end{array}
\right]
\left(
\begin{array}{c}
 A \\
 B
\end{array}
\right),
\end{equation}
which can be expressed in compact notation as
\begin{equation}
\label{Eq_2}
i \frac{d\ket{\Psi}}{dt} =\hat{H}'(t)\ket{\Psi},
\end{equation}
with $\ket{\Psi}=(A \; B)^T$ and
\begin{equation}
\label{H}
\hat{H}'(t)
= -\frac{J}{2} \hat{\mathbb{I}}_2+\frac{\Omega (t)}{2} \hat{\sigma}_x + \frac{J}{2} \hat{\sigma}_z,
\end{equation}
$\hat{\mathbb{I}}_2$ being the $2\times 2$ identity matrix. It is worth mentioning that the reduction to a two-level system is possible because the three-level system is not controllable, as becomes obvious from transformation (\ref{A_B_Gamma}), where variable $C$ is constant and cannot be altered by the control.
% \subsection{More than an angle}

% When 

% \begin{subequations}
% \label{c1_c0_final}
%     \begin{eqnarray}
%     u_x &=& \Omega (t) \cos (\omega t + \varphi)\\
%     u_y &=& \Omega (t) \sin (\omega t + \varphi)
%     \end{eqnarray}
% \end{subequations}

% And by using Eqs. (\ref{alpha_to_c}) and Eq. (\ref{time_evolution_hamiltonian}) we find:

% \begin{equation}
% i \dot{\mathbf{c}} = \left[
% \begin{array}{ccc}
%  0 & \frac{\Omega (t) e^{-i \varphi }}{2 \sqrt{2}} & 0 \\
%  \frac{\Omega (t) e^{i \varphi }}{2 \sqrt{2}} & -J  & \frac{\Omega (t) e^{-i \varphi }}{2 \sqrt{2}} \\
%  0 & \frac{\Omega (t) e^{i \varphi }}{2 \sqrt{2}} & 0 \\
% \end{array}
% \right] \mathbf{c}
% \end{equation}

% \subsection{Continue Here}

Starting from the spin-down state $\ket{\psi_0}$ the initial conditions are $a_0(0)=1, a_1(0)=a_2(0)=0$, which are the same for $c_i$, we thus get $A(0)=1/\sqrt{2},B(0)=0,C(0)=-1/\sqrt{2}$. Note the normalization $\|\ket{\Psi}\|=1/\sqrt{2}$ for the effective two-level system (\ref{two_level}).
The goal of a charging protocol $\Omega(t)$ applied during the time interval $[0 \; T]$ is to maximize the stored energy at the final time. Observe that transformation (\ref{alpha_to_c}) preserves the populations of states, thus expression (\ref{E_stored}) for the stored energy remains valid if we replace $a_i$ with $c_i$.
Using Eq. (\ref{A_B_Gamma}) and the fact that 
$C(t)$ is constant, thus $C(T)=C(0)=-1/\sqrt{2}$, we get
\begin{subequations}
\begin{eqnarray}
    c_0(T) &=& \frac{1}{\sqrt{2}}\left [A(T)+\frac{1}{\sqrt{2}}\right], \\
    c_2(T) &=& \frac{1}{\sqrt{2}}\left [A(T)-\frac{1}{\sqrt{2}}\right],
\end{eqnarray}
\end{subequations}
thus the quantity to be maximized can also be expressed as 
\begin{equation}
\label{A_Stored}
\frac{\Delta E}{\Omega_z} = \chi \left( \left|A(T)\right|^2 -\frac{1}{2} \right) - \frac{\Re(A(T))}{\sqrt{2}}+\frac{1}{2},
\end{equation}
where $\Re()$ denotes the real part. Observe that if the control $\Omega(t)$ is selected such that at the final time $t=T$ we have $A(T)=-1/\sqrt{2}$, which also implies that $B(T)=0$, then $|c_2(T)|^2=1$ (all the population is transferred to the spin-up state), the stored energy is maximized and complete charging is achieved. 

In order to gain some intuition regarding the charging process, let us consider a constant control $\Omega(t)=\Omega_0$ applied for the whole time interval $T$. The corresponding propagator is
\begin{equation}
U=e^{\frac{i}{2}JT}U^T_{\Omega_0}
\end{equation}
with
\begin{equation}
U^T_{\Omega_0}=e^{-\frac{i}{2}\omega T(n_x\sigma_x+n_z\sigma_z)}
\end{equation}
and
\begin{subequations}
\begin{eqnarray}
\label{parameters}
\omega&=&\sqrt{\Omega_0^2+J^2},\\
n_x&=& \frac{\Omega_0}{\omega},\\
n_z&=&\frac{J}{\omega},
\end{eqnarray}
\end{subequations}
driving the two level system to the final state
\begin{equation}
\label{final_state}
\ket{\Psi(T)} =
\left(
\begin{array}{c}
 A(T) \\
 B(T)
\end{array}
\right)
=\frac{e^{\frac{i}{2}JT}}{\sqrt{2}}\left(
\begin{array}{c}
\cos{\frac{\omega T}{2}}-in_z\sin{\frac{\omega T}{2}}\\
 -in_x\sin{\frac{\omega T}{2}}
\end{array}
\right).
\end{equation}
For full charging it is necessary that $B(T)=0\Rightarrow\sin{(\omega T/2)}=0$, thus $\omega T=2m\pi$ with $m=1, 2,\ldots$ and $\cos{(\omega T/2)}=(-1)^{m}$. The additional requirement $A(T)=-1/\sqrt{2}$ leads to the condition $e^{\frac{i}{2}JT}=(-1)^{m-1}=\pm 1$. The shorter duration for which this relation is satisfied is $T=2\pi/J$, and if we choose the smallest integer $m=2$ we can find from $\omega T=4\pi$ the smallest possible value $\Omega_0^{\text{min}}$.
Thus, the shortest duration and the smallest value of
a constant pulse achieving full charging are
\begin{equation}
\label{time_constant}
T = \frac{2\pi}{J}, \quad \Omega_0^{\text{min}} = \sqrt{3}J \approx 1.732 J.
\end{equation}
It is interesting to note from Eq. (\ref{final_state}) that, if $\Omega_0$ is infinite, then a delta pulse with strength $\Omega_0 T= 2\pi$ can generate instantaneously the desired $\pi$-phase in the up-state, since $T\rightarrow 0$, and thus achieve complete charging. For large but finite $\Omega_0$, duration $T=2\pi/\Omega_0$ is also finite and the dynamic phase term $e^{\frac{i}{2}JT}$, though close to unity, prohibits the full charging of the battery. In order to break the time limit (\ref{time_constant}), we need to use protocols with more pulses.

We thus seek out protocols which in the limit of very large $\Omega_0$ can achieve complete charging faster than duration (\ref{time_constant}), with the hope that the modified versions of these protocols for smaller $\Omega_0$ also break this time limit. Suppose for a moment that the control upper bound may attain very large values, so hard pulses are allowed, and consider the pulse-sequence
\begin{equation}
\label{pipi}
\pi_{x} - T - \pi_{-x}, \quad T=\frac{\pi}{J}
\end{equation}
consisting of an initial $\pi$-pulse with $x$ phase, followed by a time interval $T=\pi/J$ of free evolution $\Omega(t)=0$, and a final $\pi$-pulse with $-x$ phase. Using Eqs. (\ref{Eq_2}) and (\ref{H}), it is not hard to verify that the total propagator corresponding to this sequence is
\begin{equation}
U=e^{\frac{i}{2}JT}e^{\frac{i}{2}\pi\sigma_x}e^{-\frac{i}{2}JT\sigma_z}e^{-\frac{i}{2}\pi\sigma_x}=-\sigma_z, \quad T=\frac{\pi}{J}
\end{equation}
which drives the effective two-level system from the initial state $\ket{\Psi(0)}=(1/\sqrt{2} \quad 0)^T$ to the final state $\ket{\Psi(T)}=(-1/\sqrt{2} \quad 0)^T$, corresponding to a full charging of the quantum battery in $T=\pi/J$ units of time. Note that this duration is consistent with Eq. (\ref{Three_level}), where the excited levels are separated by frequency $J$. According to the two-level Eq. (\ref{two_level}), the first $\pi$-pulse inverts the population to the down-state, where the dynamic phase is built during the middle Off pulse, while the second $\pi$-pulse brings the population back to the up-state. The total phase acquired during the whole process is $\pi$. Note that if $\Omega(t)$ is restricted to non-negative values, then the same result can be obtained using the sequence
\begin{equation}
\label{trepipi}
(3\pi)_{x} - T - \pi_{x}, \quad T=\frac{\pi}{J}.
\end{equation}

In this article we are interested in the fast charging of the spin-pair quantum battery, we thus consider that the upper bound of the control function $\Omega(t)$ satisfies $\Omega_0>\Omega_0^{\text{min}}=\sqrt{3}J$. In this case, the full charging time is obviously shorter than $2\pi/J$. Thus, we restrict our analysis to durations $T<2\pi/J$ and ask what is the optimal $\Omega(t)$ in the interval $[0 \; T]$ maximizing the stored energy (\ref{A_Stored}) at the final time. In the following section we study this optimal control problem for the case where the control is restricted to be non-negative,  
\begin{equation}
\label{positive}
0 \leq \Omega(t) \leq \Omega_0,
\end{equation}
and the case where it is allowed to take values in a symmetric domain around zero,
\begin{equation}
\label{symmetric}
-\Omega_0 \leq \Omega(t) \leq \Omega_0,
\end{equation}
where note that negative $\Omega(t)$ corresponds to a total transverse field pointing in the opposite direction. The existence of a solution for this optimal control problem is guaranteed since $T$ is fixed, the control set is compact and the set of velocities in system (\ref{two_level}) is convex, thus the reachable set in finite time is compact and each point on its boundary can be reached by an optimal trajectory \cite{Boscain21}.
From the previous analysis, we expect for shorter durations a single constant pulse to be optimal, while for $T>\pi/J$ to come into play modified versions of pulse-sequences (\ref{pipi}) and (\ref{trepipi}) for finite $\Omega_0$. We also study for both cases the problem of full charging in minimum time.

\section{Derivation of optimal charging protocols using control theory}
\label{sec:Optimal_Solution}

We use the optimal control formalism as adapted to the study of quantum systems \cite{SugnyReview}. For the problem under consideration the quantity to be maximized, given in Eq. (\ref{A_Stored}), depends only on the terminal state and there is no running cost. If $\ket{\lambda(t)}=[\lambda_A(t), \lambda_B(t)]^T$ denotes the adjoint ket state, then the control Hamiltonian is
\begin{eqnarray}
\label{H_c}
H_c &=& \Re{(\lambda_A)}\Re{(\dot{A})}+\Im{(\lambda_A)}\Im{(\dot{A})}+\Re{(\lambda_B)}\Re{(\dot{B})}+\Im{(\lambda_B)}\Im{(\dot{B})} \nonumber \\
    &=& \Re{[\braket{\lambda|\dot{\Psi}}]} \nonumber \\
    &=& \Re{\left[-i \braket{\lambda|H'|\Psi}\right]} \nonumber \\
    &=& c'+ \phi_x(t) \Omega (t) + \phi_z(t) J,
\end{eqnarray}
where
\begin{subequations}
\label{im_phi_set}
    \begin{eqnarray}
    \phi_x &=& - \frac{1}{2} \Re{[i\braket{\lambda| \hat{\sigma}_x |\Psi}]}, \\
    \phi_y &=& - \frac{1}{2} \Re{[i\braket{\lambda| \hat{\sigma}_y |\Psi}]}, \\
    \phi_z &=& - \frac{1}{2} \Re{[i\braket{\lambda| \hat{\sigma}_z |\Psi}]},
    \end{eqnarray}
\end{subequations}
and
\begin{equation}
\label{cprime}
c'= \frac{J}{2}\Re{[i\braket{\lambda | \Psi}]}.   
\end{equation}
Using Hamilton's equations for the adjoint variables
\begin{eqnarray}
\label{costates}
    \Re(\dot{\lambda}_A) &=& -\frac{\partial H_c}{\partial\Re(A)}, \nonumber \\
    \Im(\dot{\lambda}_A) &=& -\frac{\partial H_c}{\partial\Im(A)}, \nonumber \\
    \Re(\dot{\lambda}_B) &=& -\frac{\partial H_c}{\partial\Re(B)}, \nonumber \\
    \Im(\dot{\lambda}_B) &=& -\frac{\partial H_c}{\partial\Im(B)},   
\end{eqnarray}
it is not hard to verify that the adjoint state satisfies the same equation as state $\ket{\Psi}$ \cite{SugnyReview}
\begin{equation}
\label{lambda}
    i \frac{d}{dt}\ket{\lambda} = H' \ket{\lambda} \rightarrow i \frac{d}{dt}\bra{\lambda} = -\bra{\lambda} H',
\end{equation}
where in the derivation of the equation for the bra $\bra{\lambda}$ we also used that $H'$ is Hermitian.
The adjoint variables also satisfy the terminal conditions
\begin{subequations}
\label{terminal_costates}
\begin{eqnarray}
    \Re\left[\lambda_A(T)\right] &=& \frac{\partial \left(\frac{\Delta E(T)}{\Omega_z}\right)}{\partial \Re[A(T)]} = 2 \chi \Re\left[A(T)\right] - \frac{1}{\sqrt{2}}, \\
    \Im\left[\lambda_A(T)\right] &=& \frac{\partial \left(\frac{\Delta E(T)}{\Omega_z}\right)}{\partial \Im[A(T)]} = 2 \chi \Im\left[A(T)\right], \\
    \Re\left[\lambda_B(T)\right] &=& \frac{\partial \left(\frac{\Delta E(T)}{\Omega_z}\right)}{\partial \Re[B(T)]} = 0, \\
    \Im\left[\lambda_B(T)\right] &=& \frac{\partial \left(\frac{\Delta E(T)}{\Omega_z}\right)}{\partial \Im[B(T)]} = 0, 
\end{eqnarray}
\end{subequations}
which can be expressed in compact form as
\begin{equation}
    \label{final_lambda}
    \ket{\lambda(T)} = \left(
    \begin{array}{c}
        2 \chi A(T) - \frac{1}{\sqrt{2}} \\
        0
    \end{array} \right).
\end{equation}
From Eqs. (\ref{Eq_2}) and (\ref{lambda}) it can be easily shown that the term $c'$ in $H_c$, defined in Eq. (\ref{cprime}), is constant. Since the problem is stationary, the control Hamiltonian $H_c$ is also constant. Putting these together we conclude that the following quantity is constant along the optimal trajectory,
\begin{equation}
\label{control_hamiltonian}
\phi_x(t) \Omega (t) + \phi_z(t) J = c.
\end{equation}

According to Pontryagin's Maximum Principle \cite{Pontryagin62,Heinz12}, the optimal control $\Omega^*(t)$ is selected to maximize the control Hamiltonian for almost all times (except possibly a measure zero set). Since the control Hamiltonian is linear in the bounded variable $\Omega$, the optimal pulse-sequence is determined by the switching function $\phi_x$ multiplying the control $\Omega(t)$. Specifically, $\Omega^*(t)=\Omega_0$ for $\phi_x>0$, while $\Omega^*(t)$ is set to the lower control bound ($0$ or $-\Omega_0$) for $\phi_x<0$. The Maximum Principle provides no information about the optimal control for finite time intervals where the switching function is zero. In such cases the control is called singular and is determined from the requirements $\phi_x=\dot{\phi}_x=\ddot{\phi}_x=\ldots=0$. It becomes obvious that, in order to find the optimal $\Omega^*(t)$, it is necessary to track the time evolution of $\phi_x$. From definition (\ref{im_phi_set}) and Eqs. (\ref{Eq_2}), (\ref{lambda}), we find that the components of $\vec{\phi}=(\phi_x, \phi_y, \phi_z)^T$ obey the equations
\begin{subequations}
\label{phi_TE}
    \begin{eqnarray}
    \dot{\phi}_x &=& -J \phi_y, \label{phix}\\
    \dot{\phi}_y &=& J \phi_x - \Omega(t) \phi_z, \label{phiy} \\
    \dot{\phi}_z &=& \Omega(t) \phi_y,
    \end{eqnarray}
\end{subequations}
which can be expressed in compact form as
\begin{equation}
    \vec{\phi} = \vec{B}(t)\times\vec{\phi}
\end{equation}
and describe rotations around the total field $\vec{B}(t)=(\Omega(t), 0, J)^T$ of the effective two-level system.

We next show that singular arcs may be part of the optimal trajectory and find the corresponding singular control. If $\phi_x=0$ for a finite time interval, then also $\dot{\phi}_x=0$ and Eq. (\ref{phix}) leads to $\phi_y=0$ for $J\neq 0$. Then also $\dot{\phi}_y=0$ and Eq.(\ref{phi_TE}b) leads to $\Omega(t) \phi_z = 0$. As we shall see in Sec. \ref{sec:full}, $\phi_z=0$ on the singular arc is encountered only for the case of full charging, with duration $T$ free. Aside this situation which is discussed there, for all the other cases with fixed $T$ it is $\phi_z \neq 0$ and the control on the singular arc is $\Omega_s(t)=0$.

Now let us consider the case where the control $\Omega(t)$ is restricted as in Eq. (\ref{positive}). According to the above analysis, the optimal control can only take the values $0$ and $\Omega_0$. From Eq. (\ref{two_level}) we see that a pulse $\Omega(t)=0$ does not change $A(t)$ and the stored energy, thus the simplest non-trivial pulse-sequence beyond a constant pulse is
\begin{equation}
\label{seq_I}
\Omega_0-0-\Omega_0 \quad\quad  \text{(Pulse-sequence I)}.
\end{equation}
One question that immediately arises is the nature of the middle Off pulse, whether it is a bang or a singular pulse. During the Off pulse $\Omega(t)=0$ and Eqs. (\ref{phi_TE}) give
\begin{equation}
\label{phi_x_TE_off}
\ddot{\phi}_x = -J^2 \phi_x.
\end{equation}
If $t=0$ denotes the start of the Off pulse, then $\phi_x(0)=0$
and the solution of the above equation is
\begin{equation}
\label{phi_x_off}
\phi_x(t) = \frac{\dot{\phi}_x(0) \sin (J  t)}{J }.
\end{equation}
If the Off pulse is a bang pulse, then $\dot{\phi}_x(0)\neq 0$ and the pulse lasts until the switching function $\phi_x$ becomes zero again, 
for $\tau_{\text{off}}=\pi/J$. On the other hand, if the Off pulse is a singular pulse, then $\dot{\phi}_x(0)= 0$ and the pulse duration is undetermined. For a fixed total duration $T$ of the pulse-sequence, we see that the first choice leaves only one free parameter, for example the duration of the first On pulse, to be optimized for the maximization of the stored energy, while the second choice offers two free optimization parameters, for example the durations of the first two pulses. From this we conclude that the middle pulse should be actually singular. A pulse-sequence with two or more Off bangs is also rejected, since its duration would be longer than $2\pi/J$. From this analysis and the discussion in the previous section becomes clear the central role of bang-singular-bang sequence for stored energy maximization. The first bang pulse moves the Bloch vector from the north pole towards the south pole, the intermediate singular pulse builds the dynamic phase, and the final bang pulse exploits the accumulated phase by bringing the Bloch vector again towards the north pole. Guided by the above reasoning, we also consider for the case where $\Omega(t)$ is restricted as in Eq. (\ref{symmetric}) and the optimal control can only take the values $0$ and $\pm\Omega_0$, the bang-singular-bang sequence
\begin{equation}
\label{seq_II}
\Omega_0-0-(-\Omega_0) \quad\quad  \text{(Pulse-sequence II)},
\end{equation}
where note that the same effect is obtained if the sequence starts with the negative bang pulse and ends with the positive one. In the next section we find the optimal timings for the pulses in the bang-singular-bang charging protocol, for both cases I and II. 

We close this section by pointing out that the possibility of singular control as part of the optimal pulse-sequence in a two-level system has been reported in the early works \cite{Boscain06,Mason08}. There, the problem of population inversion was extensively studied and it was shown that the optimal solution has actually the bang-bang form, without singular arcs. This is exactly the theory where is based the recent work \cite{PhysRevA.107.032218} regarding the fast charging of a single-qubit quantum battery, as well as the work \cite{PhysRevA.108.062425} regarding the fast generation of a uniform superposition in a qubit. But in the present work we study a two-qubit quantum battery whose time evolution can be mapped to a single-qubit equation, as explained in Sec. \ref{sec:II}. This imposes different requirements on the final state of this ``artificial" qubit. For example, in the case of full charging, its Bloch vector should return to the north pole while the qubit acquires a global $\pi$-phase. As a consequence, the optimal pulse-sequence now includes a singular pulse. The bang-singular-bang solution is also consistent
with the theory of Ref. \cite{Khaneja01}, developed for the case where hard pulses are allowed while here we consider the control upper bound to have finite values. We note that for the case where the lower control bound is zero, bang-bang sequences can be excluded using the arguments of the previous paragraph. When the lower bound is $-\Omega_0$, extensive numerical simulations using the optimal control solver BOCOP \cite{bocop} indicate that the bang-singular-bang pulse-sequence is optimal for the values $\Omega_0>\sqrt{3}J$ considered here, in consistency with Ref. \cite{Khaneja01} where hard pulses are permitted. 
All the above observations, although they do not constitute a strict mathematical proof, provide good indications regarding the optimality of the bang-singular-bang pulse-sequence.
Actually, often BOCOP gives solutions with more singular arcs, but the stored energy achieved is the same with that obtained by the simpler bang-singular-bang sequence. This is possibly related to the appearance of a complicated cut locus in the minimum time problem on a qubit, see Refs. \cite{Boscain06,Mason08}, implying that multiple candidates for optimality might intersect, in which case strictly proving the optimality of a pulse-sequence is not an easy task.
We finally point out that singular pulses are also encountered in the fast preparation of a maximally entangled state in a similar system composed of coupled superconducting gmon qubits \cite{PhysRevA.97.062343}.

\section{Bang-Singular-Bang charging protocol}

\label{sec:bsb}

For a bang-singular-bang candidate optimal pulse-sequence we find the durations $\tau_i, i=1, 2,3$, of the pulses. We use that on the singular arc it is $\phi_x=\phi_y=0$, and specifically at the second switching time $t=\tau_1+\tau_2$,
\begin{eqnarray}
\label{phi_xy}
    \phi_{x,y}(\tau_1+\tau_2) &=& \Re{\left[ - \frac{i}{2} \braket{\lambda(\tau_1+\tau_2)|\sigma_{x,y}|\Psi(\tau_1+\tau_2)} \right]} \nonumber \\
                              &=& 0.
\end{eqnarray}
We find $\ket{\lambda(\tau_1+\tau_2)}$ using terminal condition (\ref{final_lambda}). The final value $A(T)$ needed there can be found from Eq. (\ref{Eq_2}) for the corresponding pulse sequence,
\begin{equation}
\label{bsb_final}
\ket{\Psi(T)} =
\left(
\begin{array}{c}
 A(T) \\
 B(T)
\end{array}
\right)
=\frac{e^{\frac{i}{2}JT}}{\sqrt{2}}U_3U_2U_1
\left(
\begin{array}{c}
1\\
0
\end{array}
\right),
\end{equation}
where $U_1=U_{\Omega_0}^{\tau_1}$, $U_2=U_0^{\tau_2}$, $U_3=U_{\pm\Omega_0}^{\tau_3}$, and the plus sign corresponds to the case where $\Omega(t)$ is restricted to be non-negative (\ref{positive}), while the minus sign to the case where it can take values in the symmetric domain (\ref{symmetric}). The product of propagators in Eq. (\ref{bsb_final}) can be expressed as
\begin{eqnarray}
    U_3U_2U_1 &=& u_I \hat{\mathbb{I}}_2 + i\cdot \left( u_x \hat{\sigma}_x + u_y \hat{\sigma}_y + u_z \hat{\sigma}_z \right) \nonumber\\
    &=& \left[
\begin{array}{cc}
 u_I + i u_z & i u_x + u_y \\
 i u_x-u_y & u_I - i u_z \\
\end{array}
\right],
\end{eqnarray}
from which we get 
\begin{equation}
\label{bsb_final}
\ket{\Psi(T)} =
\left(
\begin{array}{c}
 A(T) \\
 B(T)
\end{array}
\right)
=\frac{e^{\frac{i}{2}JT}}{\sqrt{2}}
\left(
\begin{array}{c}
u_I + i u_z
\\
i u_x-u_y
\end{array}
\right),
\end{equation}
where parameters $u_I, u_z$ determining $A(T)$ are given in appendix \ref{sec:appendixB} for both pulse-sequences I and II as functions of 
\begin{eqnarray}
    \tau_s &=& \tau_1+\tau_3\quad \mbox{(sum of bang durations)},\\
    \tau_d &=& \tau_1-\tau_3\quad \mbox{(difference of bang durations)}, 
\end{eqnarray}
and the total duration $T$ (note also that the duration of the intermediate singular pulse is $\tau_2=T-\tau_s$) .
From Eq. (\ref{final_lambda}) we get
\begin{eqnarray}
    \label{final_l}
    \ket{\lambda(T)} &=& 
    \left(
    \begin{array}{c}
        \frac{2 \chi e^{\frac{i J  T}{2}} (u_I+i u_z) - 1}{\sqrt{2}} \\
        0 
    \end{array} 
    \right) \nonumber \\
                     &=& \frac{e^{\frac{i J  T}{2}}}{\sqrt{2}} \left[ 2 \chi (u_I+i u_z) - e^{\frac{-i J  T}{2}}\right] \left(
    \begin{array}{c}
        1 \\
        0 
    \end{array} \right).
\end{eqnarray}
But from Eq. (\ref{lambda}) we have
\begin{equation}
    \ket{\lambda(\tau_1+\tau_2)} = e^{\frac{-i J  \tau_3}{2}} U_3^\dag \ket{\lambda(T)}
\end{equation}
which, in combination with Eq. (\ref{final_l}), gives
\begin{equation}
    \label{lambda_bra}
    \bra{\lambda(\tau_1+\tau_2)} = \frac{e^{\frac{-i J  (\tau_1+\tau_2)}{2}}}{\sqrt{2}} \left[ 2 \chi (u_I-i u_z) - e^{\frac{i J  T}{2}}\right] \left(
    \begin{array}{cc}
        1 & 0 
    \end{array} \right) U_3.
\end{equation}

On the other hand, from Eq. (\ref{Eq_2}) we obtain
\begin{equation}
    \label{psi_tau1_tau2}
    \ket{\Psi(\tau_1+\tau_2)} = \frac{e^{\frac{i J  (\tau_1+\tau_2)}{2}}}{\sqrt{2}} U_2 U_1 \left(
    \begin{array}{c}
        1 \\
        0 
    \end{array} \right). 
\end{equation}
Using Eqs. (\ref{lambda_bra}), (\ref{psi_tau1_tau2}) in Eq. (\ref{phi_xy})
and by setting
\begin{subequations}
\begin{eqnarray} 
    \label{setting_xs_ys}
    \left(\begin{array}{cc} 1 & 0 \end{array} \right) U_3 \sigma_x U_2 U_1 \left(
    \begin{array}{c}
        1 \\
        0 
    \end{array} \right) &=& i x_I + x_z, \\\left(\begin{array}{cc} 1 & 0 \end{array} \right) U_3 \sigma_y U_2 U_1 \left(
    \begin{array}{c}
        1 \\
        0 
    \end{array} \right) &=& i y_I + y_z,
\end{eqnarray}
\end{subequations}
where parameters $x_I, x_z, y_I, y_z$ are given as functions of $\tau_s, \tau_d, T$ in appendix \ref{sec:appendixB},
we find that
\begin{subequations}
\label{phi_xy_t1t2}
\begin{eqnarray} 
    \phi_x(\tau_1+\tau_2) &=& \frac{\left(2 \chi u_I - \cos \frac{J  T}{2}\right) x_I - \left(2 \chi u_z + \sin \frac{J  T}{2}\right) x_z}{4} \nonumber \\ 
                          &=& 0, \\
    \phi_y(\tau_1+\tau_2) &=& \frac{\left(2 \chi u_I - \cos \frac{J  T}{2}\right) y_I - \left(2 \chi u_z + \sin \frac{J  T}{2}\right) y_z}{4} \nonumber \\
                          &=& 0.
\end{eqnarray}
\end{subequations}
Eqs. (\ref{phi_xy_t1t2}) form a system for the unknowns $\tau_s$ and $\tau_d$. Note that quantities $2 \chi u_I - \cos (JT/2)$ and $2 \chi u_z + \sin(JT/2)$ cannot be simultaneously zero because this leads to $|A(T)|=1/(2\sqrt{2} \chi )>1/\sqrt{2}$, since $ \chi <1/2$, which violates $\|\ket{\Psi}\|=1/\sqrt{2}$. Consequently, from the homogeneous system (\ref{phi_xy_t1t2}) we get
\begin{equation}
    \label{d_condition}
    x_I y_z = x_z y_I.
\end{equation}

If we substitute either expressions (\ref{xs_ys_equal}) or (\ref{xs_ys_inverted}) in Eq. (\ref{d_condition}) we end up with the same condition $\sin (\omega \tau_d/2)=0$, 
which leads to $\tau_d=0$ or $\tau_d=2\pi/\omega$.
To identify which value is optimal for each case, we consider the real part of $A(T)$,
\begin{equation}
\label{RealA}
    \Re{\left[A(T)\right]} = \frac{u_I \cos \frac{J T}{2}-u_z \sin \frac{J T}{2}}{\sqrt{2}}
\end{equation}
which, for stored energy maximization, should be as close as possible to its minimum value $-1/\sqrt{2}$.
Since $\pi/2<JT/2<\pi$, thus $\sin(JT/2)>0$ and $\cos(JT/2)<0$, we see that $u_z$ should be as positive as possible and $u_I$ as negative as possible. For pulse-sequence I (\ref{seq_I}), the dependence on $\tau_d$ is restricted in $u_z$, as we can observe from Eq. (\ref{u_pos}). Since $J(T-\tau_s)/2<JT/2<\pi$, thus $\sin[J(T-\tau_s)/2]>0$, from the same equation we conclude that $u_z$ is maximized for
\begin{equation}
    \label{tdn0}
    \tau_d=\frac{2\pi}{\omega},
\end{equation}
corresponding to $\cos (\omega \tau_d/2)=-1$. Using Eq. (\ref{tdn0}) in either of Eqs. (\ref{phi_xy_t1t2}), we obtain the following transcendental equation for $\tau_s$
\begin{widetext}
\begin{equation}
\label{equ_to_find_ts_equal}
    \frac{2 \chi \sin \left[\frac{J \left(T-\tau_s\right)}{2}\right] \left(n_x^2-n_z^2 \cos \frac{\omega \tau_s}{2}\right)-2 \chi n_z \sin \frac{\omega \tau_s}{2} \cos \left[\frac{J \left(T-\tau_s\right)}{2}\right]+\sin \frac{J  T}{2}}{2 \chi n_z \sin \frac{\omega \tau_s}{2} \sin \left[\frac{J \left(T-\tau_s\right)}{2}\right]-2 \chi \cos \frac{\omega \tau_s}{2} \cos \left[\frac{J \left(T-\tau_s\right)}{2}\right]+\cos \frac{J  T}{2}} = \frac{1}{n_z} \tan \frac{\omega \tau_s}{4}.
\end{equation}
For pulse-sequence II (\ref{seq_II}), the dependence on $\tau_d$ is contained only in $u_I$, which is minimized for 
\begin{equation}
    \label{td0}
    \tau_d=0,
\end{equation}
corresponding to $\cos (\omega \tau_d/2)=1$. Using Eq. (\ref{td0}) in either of Eqs. (\ref{phi_xy_t1t2}), we obtain another transcendental equation for $\tau_s$
\begin{equation}
\label{equ_to_find_ts_inverted}    
    \frac{\sin \frac{J  T}{2} - 2 \chi n_z \sin \frac{\omega \tau_s}{2} \cos \left[\frac{J \left(T-\tau_s\right)}{2}\right] - 2 \chi \cos \frac{\omega \tau_s}{2} \sin \left[\frac{J \left(T-\tau_s\right)}{2}\right]}{\cos \frac{J  T}{2} - 2 \chi \cos \left[\frac{J \left(T-\tau_s\right)}{2}\right] \left(n_z^2 \cos \frac{\omega \tau_s}{2}+n_x^2\right)+ 2 \chi n_z \sin \frac{\omega \tau_s}{2} \sin \left[\frac{J \left(T-\tau_s\right)}{2}\right]} = n_z \tan \frac{\omega \tau_s}{4}.
\end{equation}
\end{widetext}
Solving Eqs. (\ref{equ_to_find_ts_equal}), (\ref{equ_to_find_ts_inverted}) we can find the durations of individual pulses in sequences I and II, respectively. For both sequences, the same stored energy is achieved if the initial and final bang pulses are interchanged.

Note that by setting in Eq. (\ref{equ_to_find_ts_equal}) $\tau_s=\tau_d$, which translates to $\tau_3=0$, we can find the threshold duration $T$ at which pulse-sequence I appears. We get the transcendental equation
\begin{equation}
    \label{large_thr}
    2 \chi \cos \left[\frac{J}{2}  \left(T-\frac{2\pi }{\omega }\right)\right]+\cos \frac{J  T}{2} = 0. 
\end{equation}

% \begin{subequations}
% \begin{eqnarray} 
%     \label{lambda_psi}
%     \braket{\lambda(\tau_1+\tau_2)|\sigma_x|\Psi(\tau_1+\tau_2)} = \nonumber \\
%     \frac{ u_I-i u_z - e^{\frac{i J  T}{2}}}{2} \left(\begin{array}{cc} 1 & 0 \end{array} \right) U_3 \sigma_x U_2 U_1 \left(
%     \begin{array}{c}
%         1 \\
%         0 
%     \end{array} \right) = \nonumber \\
%     \frac{ u_I-i u_z - e^{\frac{i J  T}{2}}}{2} (i x_I + x_z) \\
%     \braket{\lambda(\tau_1+\tau_2)|\sigma_y|\Psi(\tau_1+\tau_2)} = \nonumber \\
%     \frac{ u_I-i u_z - e^{\frac{i J  T}{2}}}{2} \left(\begin{array}{cc} 1 & 0 \end{array} \right) U_3 \sigma_y U_2 U_1 \left(
%     \begin{array}{c}
%         1 \\
%         0 
%     \end{array} \right) = \nonumber \\
%     \frac{ u_I-i u_z - e^{\frac{i J  T}{2}}}{2} (i y_I + y_z)
% \end{eqnarray}
% \end{subequations}

% So now from Eqs. (\ref{im_phi_set}, \ref{re_phi_set}) we can find $\phi_i(t_f)$ for $i=x,y,z$

% %\begin{widetext}
% \begin{subequations}
% \label{final_phi}
%     \begin{eqnarray}
%     \phi_x(t_f) &=& \frac{u_I u_x+u_x \left(-\cos \frac{J  T}{2}\right)+u_y \sin \frac{J  T}{2}+u_y u_z}{4} \\
%     \phi_y(t_f) &=& \frac{u_I u_y+u_x \left(-\sin \frac{J  T}{2}\right)-u_y \cos \frac{J  T}{2}-u_x u_z}{4} \\
%     \phi_z(t_f) &=& \frac{u_I \left(-\sin \frac{J  T}{2}\right)-u_z \cos \frac{J  T}{2}}{4}
%     \end{eqnarray}
% \end{subequations}
% %\end{widetext}

\section{Minimum Time for Full Charging}

\label{sec:full}

\begin{figure}[h]
 \centering
 \begin{subfigure}[b]{0.4\textwidth}
    \centering\caption{}\includegraphics[width=\linewidth]{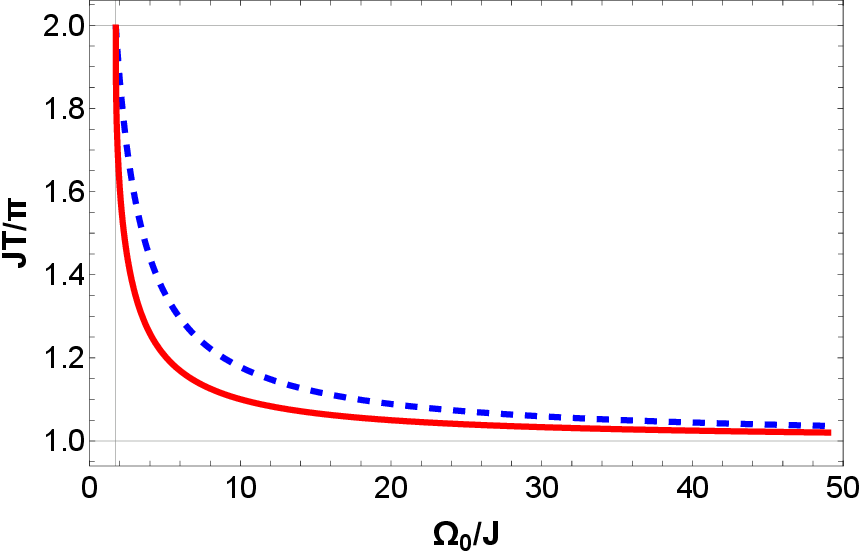}\label{fig:dur_1}
\end{subfigure}
\hspace{.2cm}
\begin{subfigure}[b]{0.4\textwidth}
    \centering\caption{}\includegraphics[width=\linewidth]{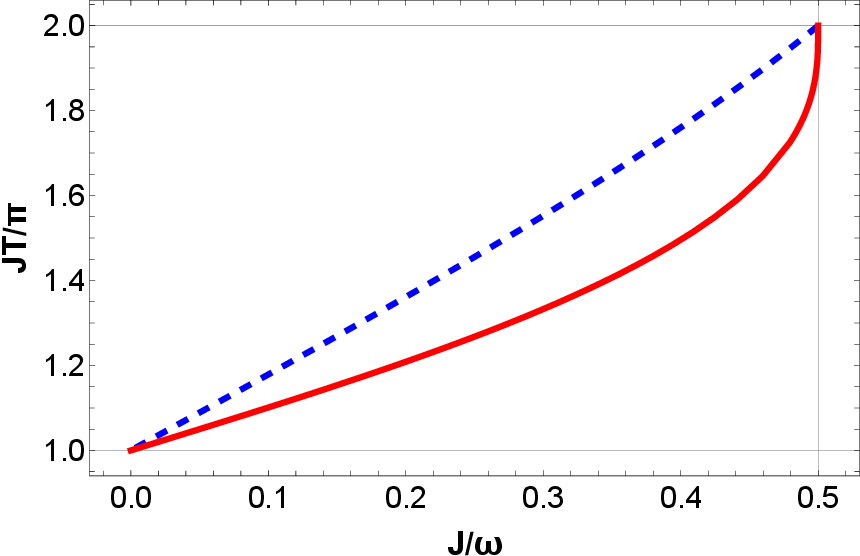}\label{fig:dur_2}
\end{subfigure}
\caption{Minimum durations for full charging, corresponding to pulse-sequences I (blue dashed curve) and II (red solid curve), as functions of (a) parameter $\Omega_0/J$ and (b) parameter $J/\omega=J/\sqrt{\Omega_0^2+J^2}$.}
\label{fig:duration}
\end{figure}

From Eq. (\ref{A_Stored}) we can see that full charging is achieved for $A(T)=-1/\sqrt{2}$, i.e. when $\Re{\left[A(T)\right]}$ attains its minimum possible value.
From Eq. (\ref{RealA}) and expressions (\ref{u_pos}), (\ref{u_sym}) for $u_I, u_z$, using also Eqs. (\ref{tdn0}) and (\ref{td0}) according to the case, we see that $\Re{[A(T)]}$
is expressed as a function of $\tau_s$,$T$. The condition for minimum $\partial \Re{\left[A(T)\right]}/\partial T=0$ gives for both cases $\sin [J(T-\tau _s/2)]=0$, which leads to the relation
\begin{equation}
\label{full_cond}
    \tau_s = 2 \left(T-\frac{\pi}{J}\right).
\end{equation}
Eqs. (\ref{equ_to_find_ts_equal}), (\ref{equ_to_find_ts_inverted}) are still valid and, using Eq. (\ref{full_cond}) to eliminate $\tau_s$, we obtain
\begin{subequations}
\label{full_T}
\begin{eqnarray}
    n_z \tan \frac{J T}{2} + \tan \left[\frac{\omega  (\pi -J T)}{2 J}\right] &=& 0, \\
    \tan \frac{J T}{2} + n_z \tan \left[\frac{\omega  (\pi -J T)}{2 J}\right] &=& 0.
\end{eqnarray}
\end{subequations}
The above transcendental equations give the minimum duration required for full charging for the two pulse-sequences. In Fig. \ref{fig:dur_1} we plot these durations as functions of parameter $\Omega_0/J$, while in Fig. \ref{fig:dur_2} we plot them as functions of parameter $J/\omega=J/\sqrt{\Omega_0^2+J^2}$, with blue dashed line for pulse-sequence I and with red solid line for pulse-sequence II. Obviously the second sequence, having a broader control range, achieves a faster full charging. 

If we make in Fig. \ref{fig:dur_2} a Taylor expansion of the curves around the lower corner point $(J/\omega,JT)=(0, \pi)$, which is approached for $\Omega_0\rightarrow\infty$, we obtain the approximation
\begin{equation}
JT\approx\pi+\alpha\frac{J}{\omega},
\end{equation}
where the slope is $\alpha=\pi$ for the red curve, while for the black curve it is the smallest positive root of the equation $\alpha/2+\cot(\alpha/2)=0$, which is $\alpha\approx 5.597$. We see that the duration of the faster pulse-sequence is increased with the smaller rate, as $\Omega_0$ is decreased and thus $J/\omega$ is increased.
A similar expansion around the upper corner point $(J/\omega,JT)=(1/2, 2\pi)$, which is approached for $\Omega_0\rightarrow\sqrt{3}J$, gives for the blue dashed curve
\begin{equation}
JT\approx2\pi-\frac{8\pi}{3}\left(\frac{1}{2}-\frac{J}{\omega}\right),
\end{equation}
while for the red curve we find
\begin{equation}
JT\approx2\pi-2(\pi)^{1/3}\left(\frac{1}{2}-\frac{J}{\omega}\right)^{1/3},
\end{equation}
by expanding the abscissa in terms of the ordinate. The duration of the faster pulse sequence drops with an infinite rate as $\Omega_0$ is increased from the value $\sqrt{3}J$ and thus $J/\omega$ is decreased from the value $1/2$.

We close this section with a very interesting observation regarding the optimal solution for full charging. From terminal condition (\ref{final_lambda}) and since $A(T)=-1/\sqrt{2}$ for full charging, we see that $\ket{\lambda(T)}=-\frac{2\chi+1}{\sqrt{2}} (1 \; 0)^T$. This condition, if used in Eqs. (\ref{im_phi_set}) along with $\ket{\Psi(T)}=-\frac{1}{\sqrt{2}} (1 \; 0)^T$, leads to $\phi_x(T)=\phi_y(T)=\phi_z(T)=0$ which, when combined with Eqs. (\ref{phi_TE}), implies that $\phi_x(t)=\phi_y(t)=\phi_z(t)=0$ along this optimal trajectory. Thus, we have found a specific quantum control example where  $\vec{\phi}=(\phi_x, \phi_y, \phi_z)^T=0$ while $\ket{\lambda}\neq 0$.

\section{Examples}
\label{sec:results}

\begin{figure}[t]
  \centering
  \begin{subfigure}[t!]{0.4\textwidth}
     \centering\caption{}\includegraphics[width=\linewidth]{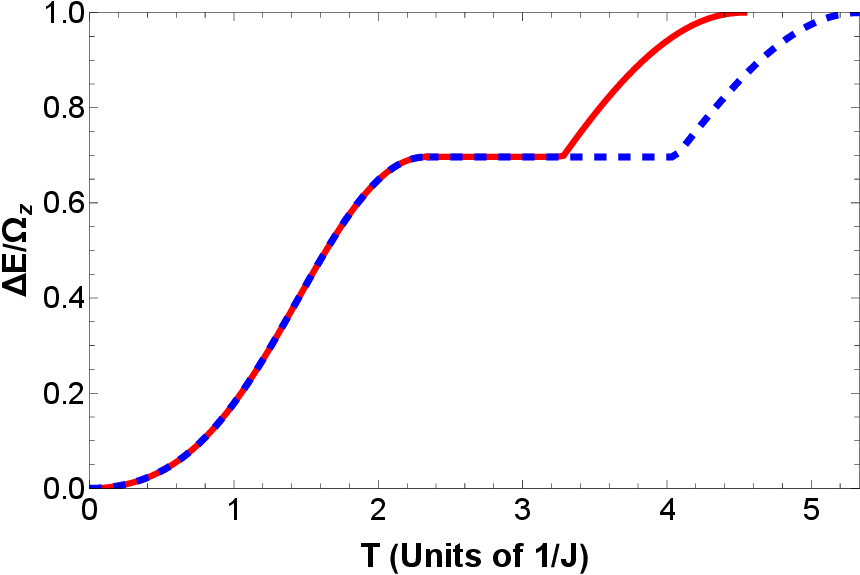}\label{fig:E_O_2_5_x_3}
 \end{subfigure} \\
 \begin{subfigure}[c]{0.4\textwidth}
     \centering\caption{}\includegraphics[width=\linewidth]{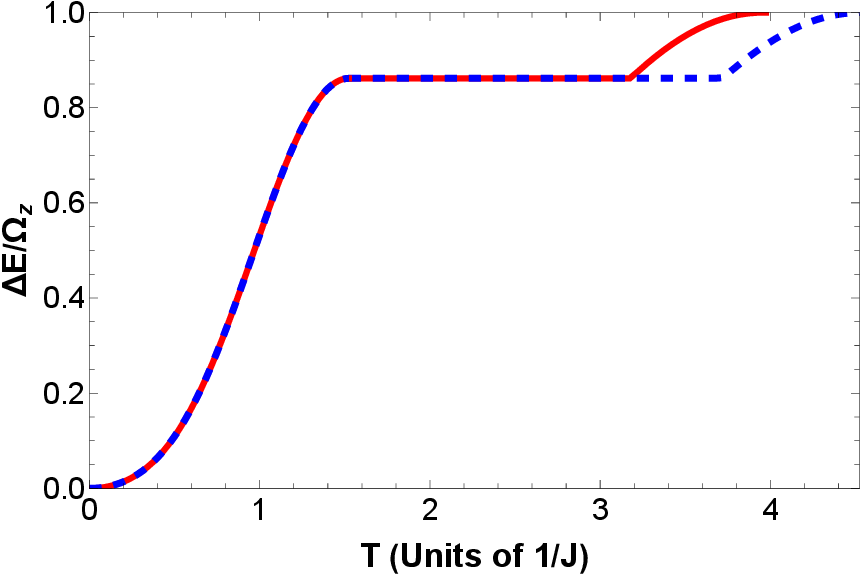}\label{fig:E_O_4_x_3}
 \end{subfigure} \\
 \begin{subfigure}[c]{0.4\textwidth}
     \centering\caption{}\includegraphics[width=\linewidth]{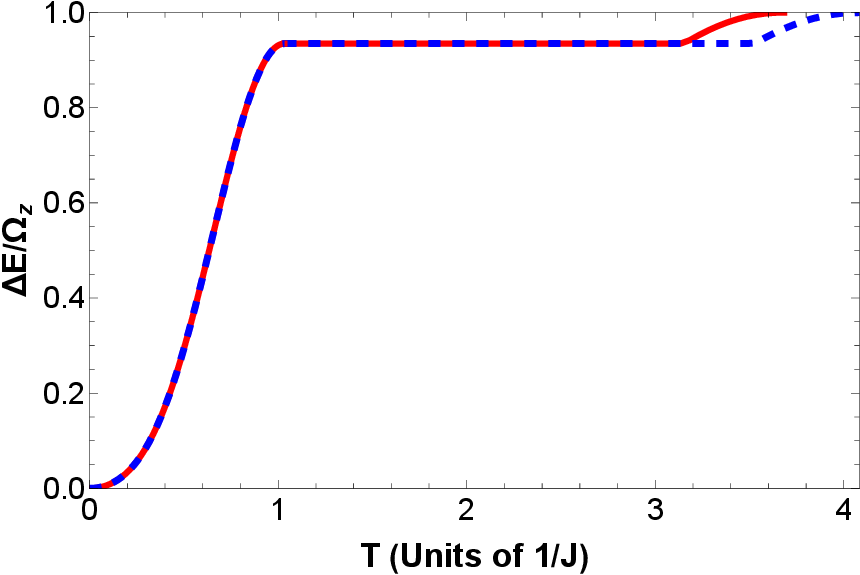}\label{fig:E_O_6_x_3}
 \end{subfigure}
 \caption{Stored energy as function of duration for the optimal pulse-sequence with $-\Omega_0 \leq \Omega(t) \leq \Omega_0$ (red solid line) and $0 \leq \Omega(t) \leq \Omega_0$ (blue dashed line) for $\chi=J/\Omega_z=1/3$ and various values of $\Omega_0/J$: (a) $\Omega_0/J=2.5$, (b) $\Omega_0/J=4$, (c) $\Omega_0/J=6$.}
 \label{fig:stored_energy}
 \end{figure}

In Fig. \ref{fig:stored_energy} we display the stored energy as a function of duration $T$ of the optimal pulse-sequence, with blue dashed line when the control is restricted as in Eq. (\ref{positive}) and with red solid line when is restricted as in Eq. (\ref{symmetric}), for various values of the control bound $\Omega_0/J = 2.5, 4, 6$ (Figs. \ref{fig:E_O_2_5_x_3}, \ref{fig:E_O_4_x_3} and \ref{fig:E_O_6_x_3}, respectively) and ratio $\chi=J/\Omega_z=1/3$. For short durations the optimal pulse-sequence is a constant pulse, consequently the achieved stored energy is the same using either domain (\ref{positive}) or (\ref{symmetric}) for the control. This region corresponds to the initial hillside in the diagrams. We observe that the stored energy achieved with a constant pulse saturates after a threshold duration equal to $2\pi/\omega$. The maximum charging
obtained with a constant pulse is higher for larger upper control bound $\Omega_0$ (from top to bottom in Fig. \ref{fig:stored_energy}), and also is attained faster. For longer durations there is an interval where any extra On pulse only degrades this performance, thus an Off pulse should follow the initial On pulse. This region corresponds to the plateau in the performance observed in all diagrams.
The bang-singular-bang pulse-sequence becomes optimal beyond another threshold duration, different for sequences I and II, with the threshold of the former given by Eq. (\ref{large_thr}) and the threshold of the latter being shorter. This region corresponds to the final hillside for both red solid and blue dashed curves, being present in all diagrams. Observe that the bang-singular-bang pulse-sequence is necessary to attain complete charging of the battery.

\begin{figure}[t]
 \centering
\includegraphics[width=\linewidth]{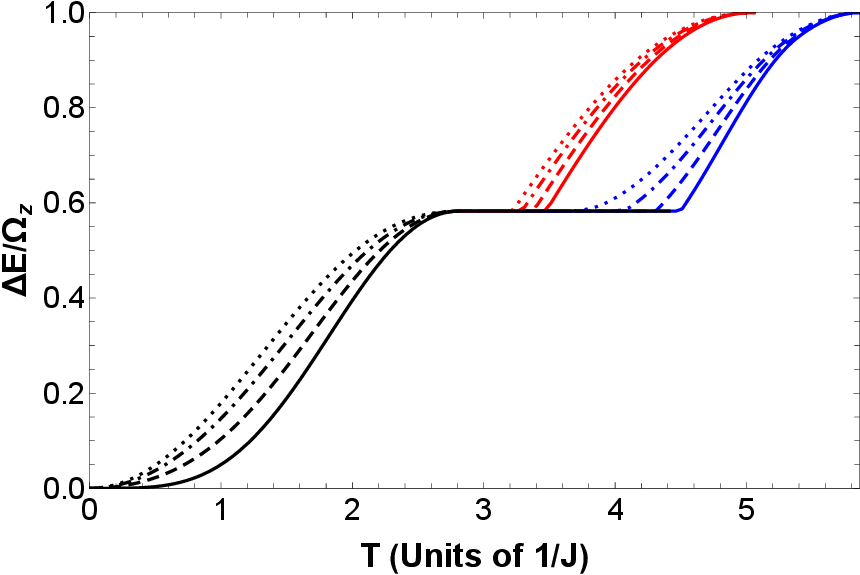}\label{fig:E_O_2_0_all_x}
\caption{Stored energy as function of duration for the optimal pulse-sequence for fixed upper control bound $\Omega_0/J=2$ and various values of the ratio $\chi=J/\Omega_z=1/2$ (solid), $1/3$ (dashed), $1/5$ (dashed-dotted) and $1/10$ (dotted). The black hillside lines correspond to the case where a single pulse is optimal, while the blue and red lines correspond to the bang-singular-bang sequence when $0 \leq \Omega(t) \leq \Omega_0$ and $-\Omega_0 \leq \Omega(t) \leq \Omega_0$, respectively.}
\label{fig:stored_energy_2}
\end{figure}

In Fig. \ref{fig:stored_energy_2} we plot the stored energy as a function of duration $T$ of the optimal pulse-sequence, for fixed $\Omega_0/J = 2$ and several values of the ratio $\chi=J/\Omega_z=1/2, 1/3, 1/5, 1/10$, denoted with solid, dashed, dashed-dotted and dotted line, respectively. We use black color for the first hillside region where a single pulse is optimal, while for the second hillside region we use blue color when the control is restricted as in Eq. (\ref{positive}), thus pulse-sequence I is optimal, and red color when the control is restricted as in Eq. (\ref{symmetric}), thus pulse-sequence II is optimal. We also use black color for the intermediate plateau where the stored energy is independent of $\chi$. Observe that the energy which can be stored within a certain duration depends on the ratio $\chi$ in general, with smaller $\chi$ (weaker interactions) corresponding to higher charging efficiency for the same duration.
But the duration needed to reach the plateau as well as for full charging are independent of $\chi$, the former being $2\pi/\omega$ while the latter is determined from transcendental Eqs. (\ref{full_T}). On the other hand, the duration for which the bang-singular-bang sequence becomes optimal and thus the charging curves leave the plateau depends on $\chi$, with smaller values of $\chi$ (weaker interactions) corresponding to earlier elevation from the plateau.

\begin{figure*}[t]
 \centering
 \vspace{-1.5cm}
 \begin{subfigure}[t!]{0.4\textwidth}
    \centering\caption{}\vspace{1.1cm}\includegraphics[width=\linewidth]{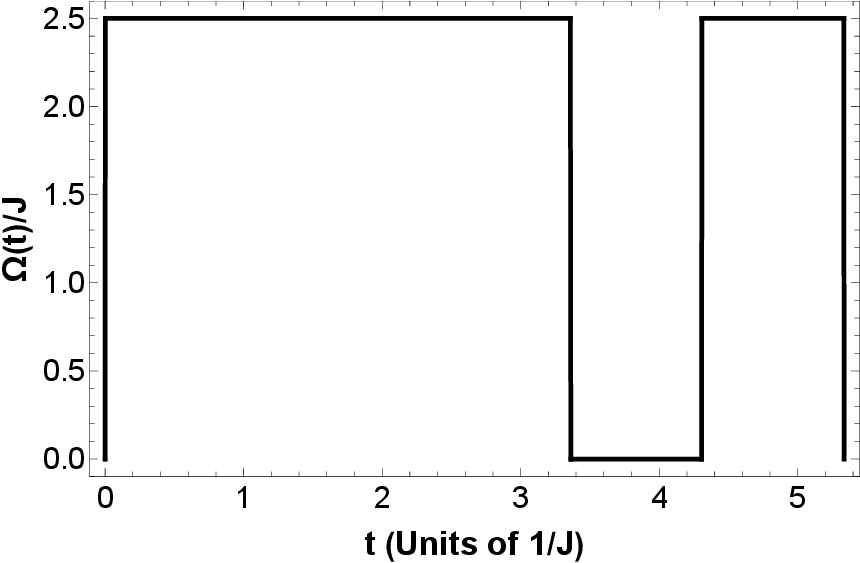}\label{fig:pcontrol_2_5}
\end{subfigure}
\begin{subfigure}[t!]{0.4\textwidth}
    \centering\vspace{1.4cm}\caption{}\includegraphics[width=\linewidth]{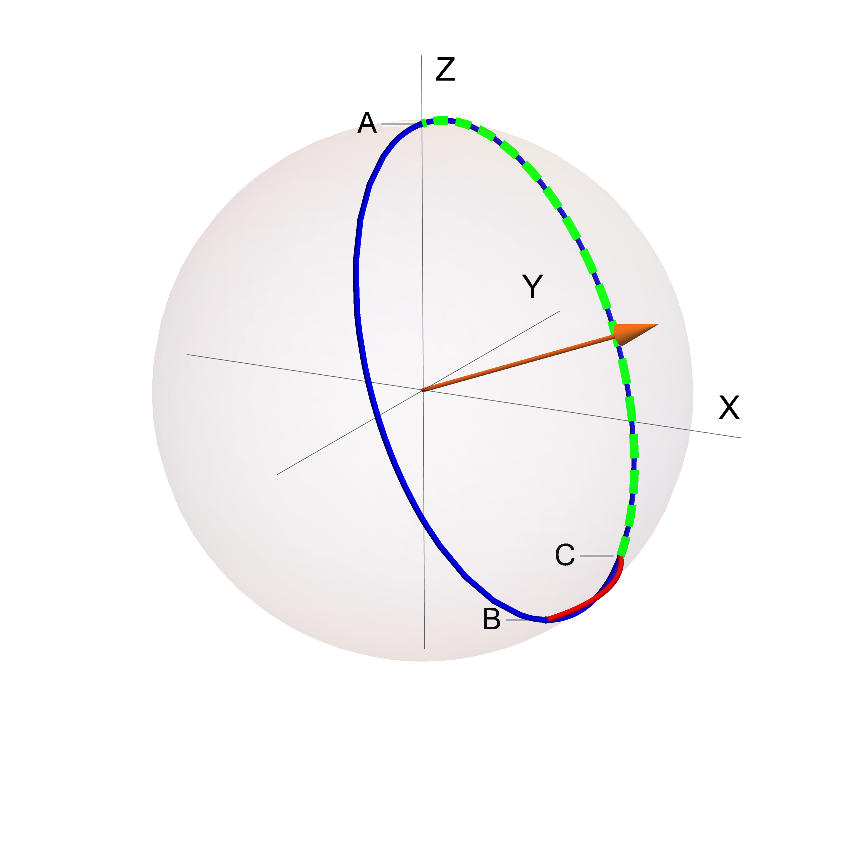}\label{fig:psphere_2_5}
\end{subfigure} \\
\vspace{-2.5cm} 
\begin{subfigure}[c]{0.4\textwidth}
    \centering\caption{}\vspace{1.1cm}\includegraphics[width=\linewidth]{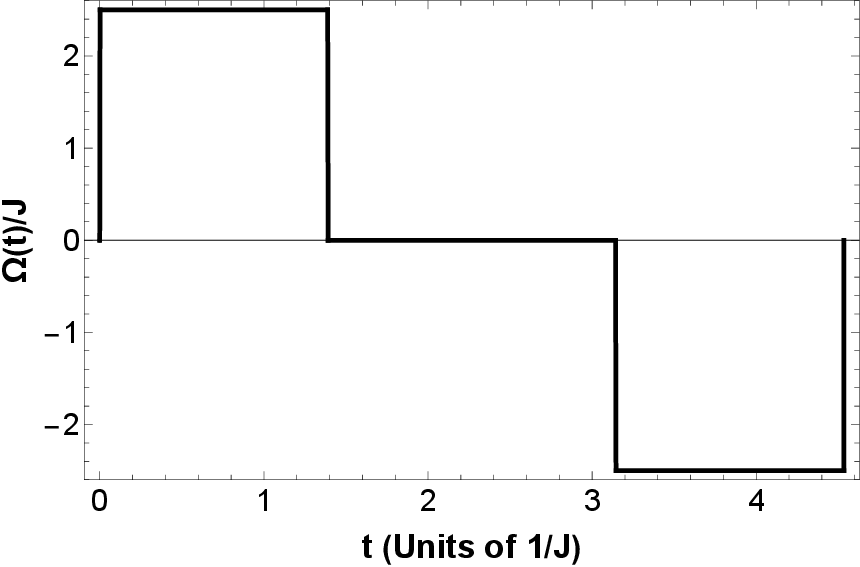}\label{fig:scontrol_2_5}
\end{subfigure}
\begin{subfigure}[c]{0.4\textwidth}
    \centering\vspace{1.4cm}\caption{}\includegraphics[width=\linewidth]{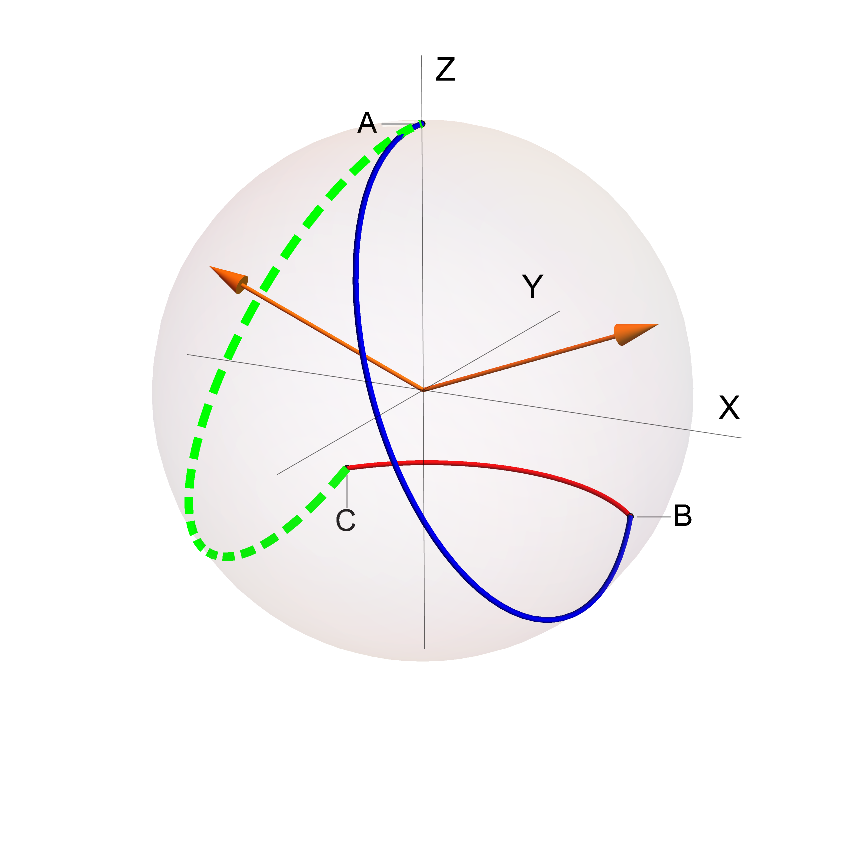}\label{fig:psphere_2_5}
\end{subfigure}
\caption{Optimal pulse-sequence and corresponding trajectory on the Bloch sphere representation for full charging in minimum time, when $\Omega_0/J=2.5, \chi=J/\Omega_z=1/3$ and (a, b) $0 \leq \Omega(t) \leq \Omega_0$, (c, d) $-\Omega_0 \leq \Omega(t) \leq \Omega_0$. Blue solid line, red solid line and green dashed line represent the segments travelled during the first bang pulse, the intermediate singular pulse and the final bang pulse, respectively.}
\label{fig:ex_2_5}
\end{figure*}

\begin{figure*}[t]
 \centering
 \vspace{-1.5cm}
 \begin{subfigure}[t!]{0.4\textwidth}
    \centering\caption{}\vspace{1.1cm}\includegraphics[width=\linewidth]{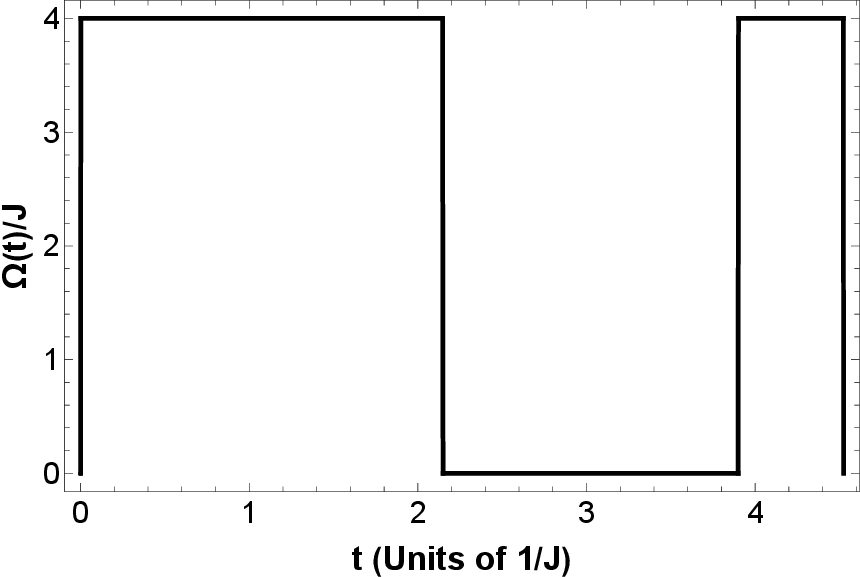}\label{fig:n_3_com_control}
\end{subfigure}
\begin{subfigure}[t!]{0.4\textwidth}
    \centering\vspace{1.4cm}\caption{}\includegraphics[width=\linewidth]{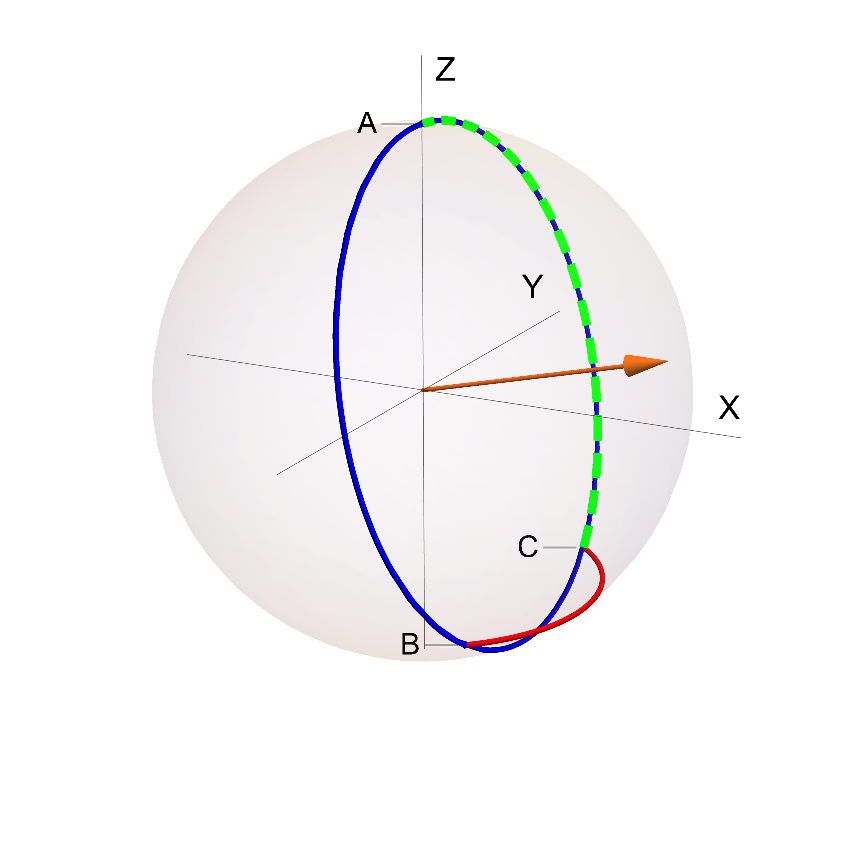}\label{fig:n_3_com_sphere}
\end{subfigure} \\
\vspace{-2.5cm} 
\begin{subfigure}[c]{0.4\textwidth}
    \centering\caption{}\vspace{1.1cm}\includegraphics[width=\linewidth]{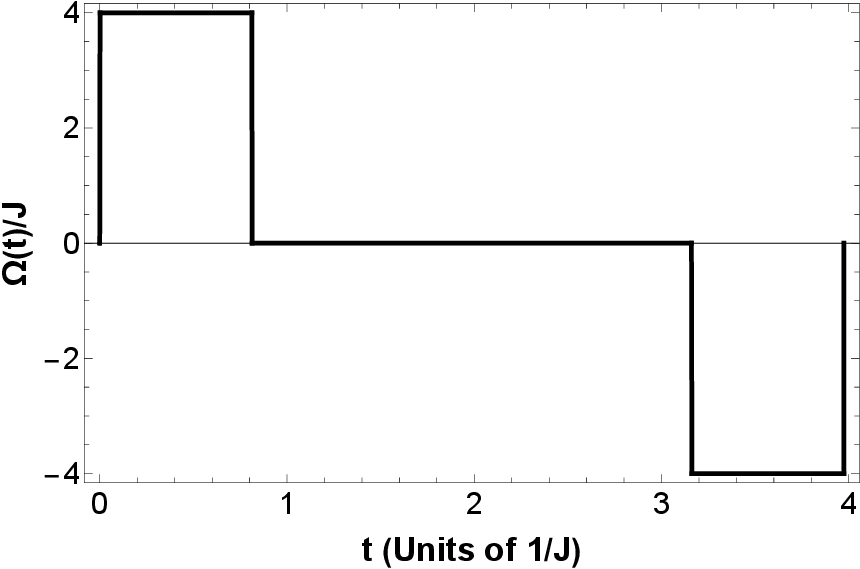}\label{fig:n_3_sym_control}
\end{subfigure}
\begin{subfigure}[c]{0.4\textwidth}
    \centering\vspace{1.4cm}\caption{}\includegraphics[width=\linewidth]{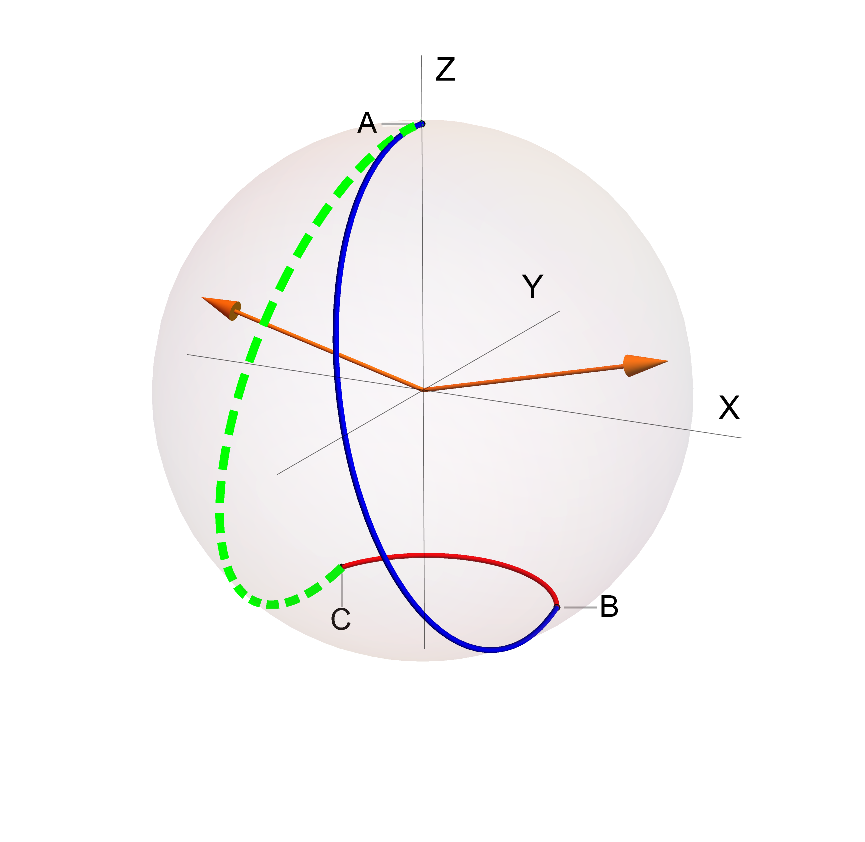}\label{fig:n_3_sym_sphere}
\end{subfigure}
\caption{Optimal pulse-sequence and corresponding trajectory on the Bloch sphere representation for full charging in minimum time, when $\Omega_0/J=4, \chi=J/\Omega_z=1/3$ and (a, b) $0 \leq \Omega(t) \leq \Omega_0$, (c, d) $-\Omega_0 \leq \Omega(t) \leq \Omega_0$. Blue solid line, red solid line and green dashed line represent the segments travelled during the first bang pulse, the intermediate singular pulse and the final bang pulse, respectively.}
\label{fig:ex_4}
\end{figure*}

\begin{figure*}[t]
 \centering
 \vspace{-1.5cm}
 \begin{subfigure}[t!]{0.4\textwidth}
    \centering\caption{}\vspace{1.1cm}\includegraphics[width=\linewidth]{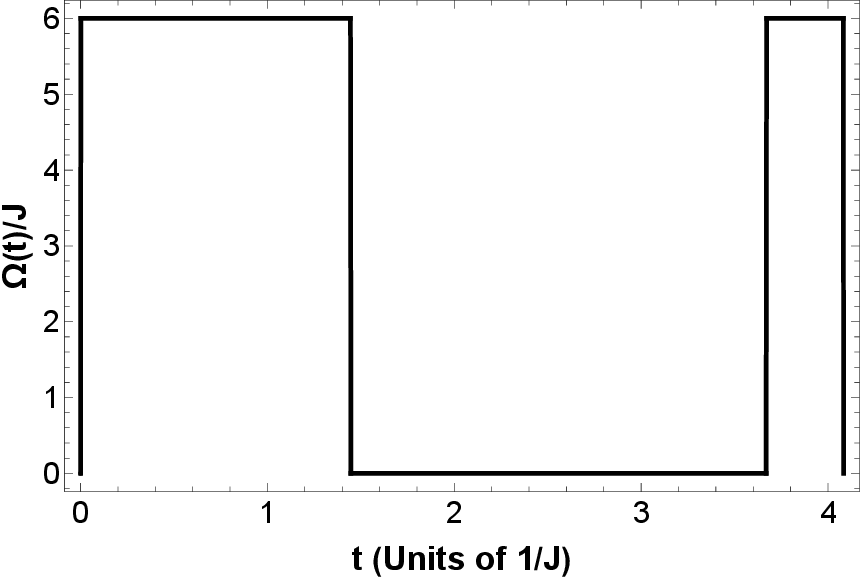}\label{fig:n_3_com_control}
\end{subfigure}
\begin{subfigure}[t!]{0.4\textwidth}
    \centering\vspace{1.4cm}\caption{}\includegraphics[width=\linewidth]{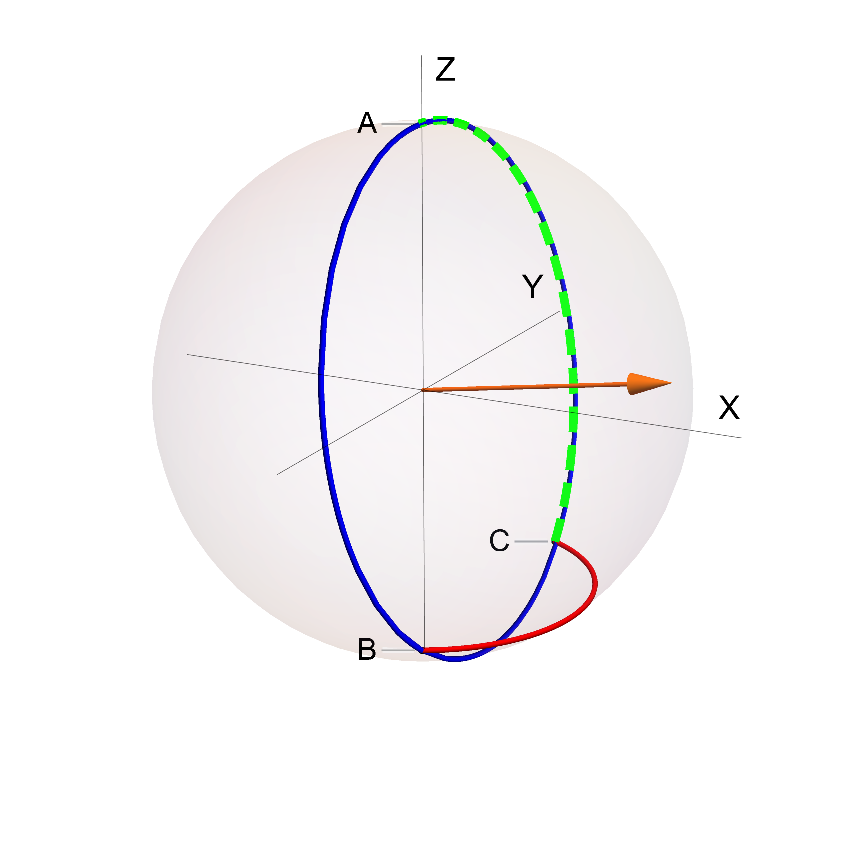}\label{fig:n_3_com_sphere}
\end{subfigure} \\
\vspace{-2.5cm} 
\begin{subfigure}[c]{0.4\textwidth}
    \centering\caption{}\vspace{1.1cm}\includegraphics[width=\linewidth]{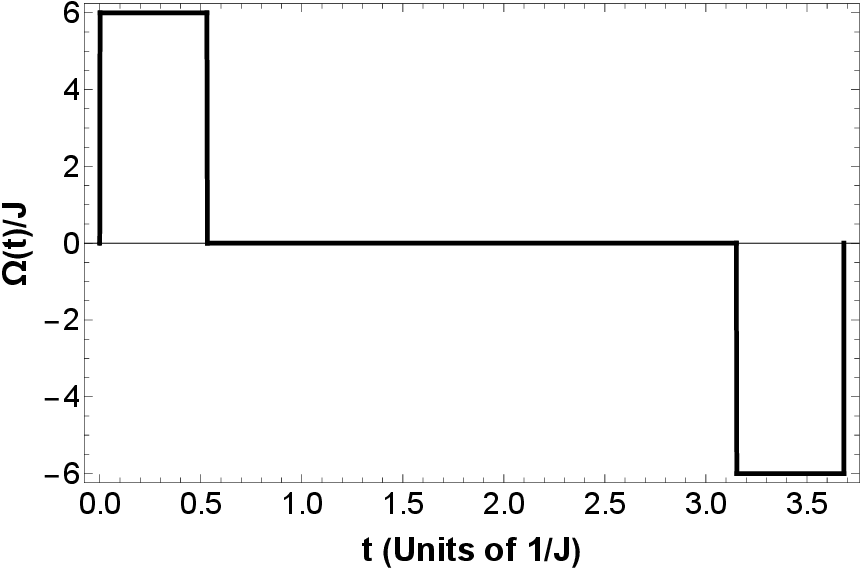}\label{fig:n_3_sym_control}
\end{subfigure}
\begin{subfigure}[c]{0.4\textwidth}
    \centering\vspace{1.4cm}\caption{}\includegraphics[width=\linewidth]{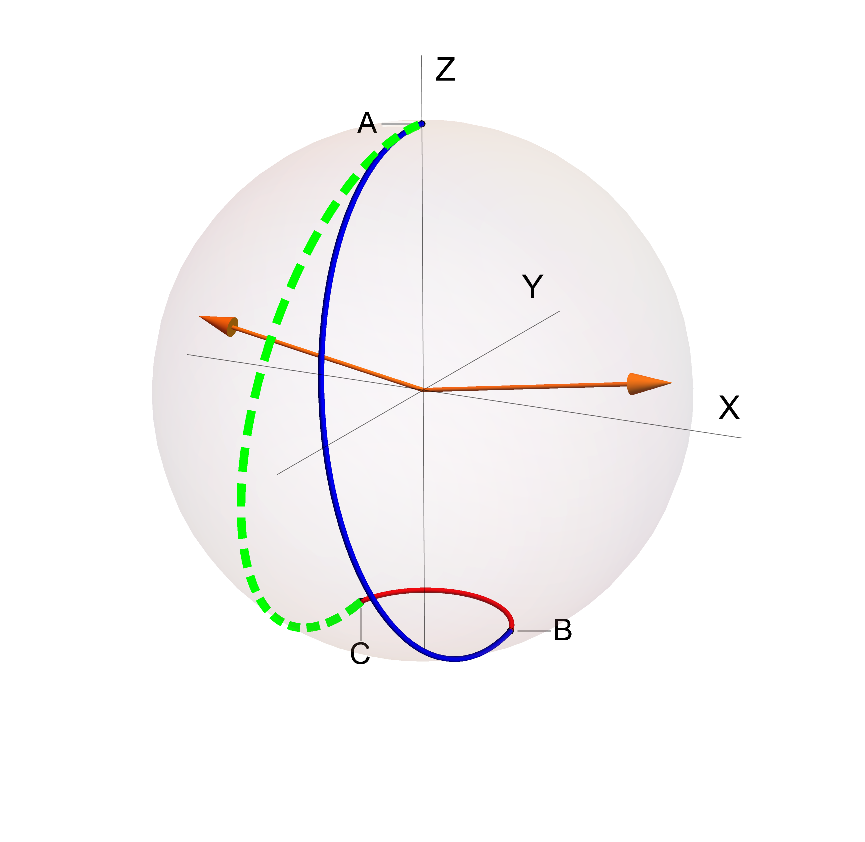}\label{fig:n_3_sym_sphere}
\end{subfigure}
\caption{Optimal pulse-sequence and corresponding trajectory on the Bloch sphere representation for full charging in minimum time, when $\Omega_0/J=6, \chi=J/\Omega_z=1/3$ and (a, b) $0 \leq \Omega(t) \leq \Omega_0$, (c, d) $-\Omega_0 \leq \Omega(t) \leq \Omega_0$. Blue solid line, red solid line and green dashed line represent the segments travelled during the first bang pulse, the intermediate singular pulse and the final bang pulse, respectively.}
\label{fig:ex_6}
\end{figure*}

In Figs. \ref{fig:ex_2_5}-\ref{fig:ex_6} we display optimal controls and the corresponding trajectories of the effective two-level system on the Bloch sphere for full charging, with upper control bound $\Omega_0/J = 2.5, 4, 6$, respectively, and fixed $\chi=J/\Omega_z=1/3$. In each figure, the first row corresponds to pulse-sequence I and the second row to pulse-sequence II. The Bloch vector points initially to the north pole and during the first bang pulse travels along the blue solid arc. For pulse-sequence I the Bloch vector completes a full rotation, thus it returns back to point A, and then continues until point B is reached. During the singular Off pulse the Bloch vector is rotated around the $z$-axis on the red solid arc, to point C. The final bang pulse brings the Bloch vector back to the north pole along the green dashed arc, after having acquired the desired $\pi$-phase. During the two bang pulses the Bloch vector is rotated around the same total effective field. For pulse-sequence II, the Bloch vector follows trajectory ABCA while the rotations corresponding to the initial and final bang pulses take place around symmetric variants of the total field. This extra rotation axis is the reason why the second sequence achieves full charging in less time and with less pulse energy than the first one. Observe that for increasing $\Omega_0/J$ the bang pulses are faster while the singular Off pulse, during which the dynamic phase is built, occupies a larger portion of the optimal pulse-sequence.

\begin{figure}[h]
 \centering
 \begin{subfigure}[b]{0.4\textwidth}
    \centering\caption{}\includegraphics[width=\linewidth]{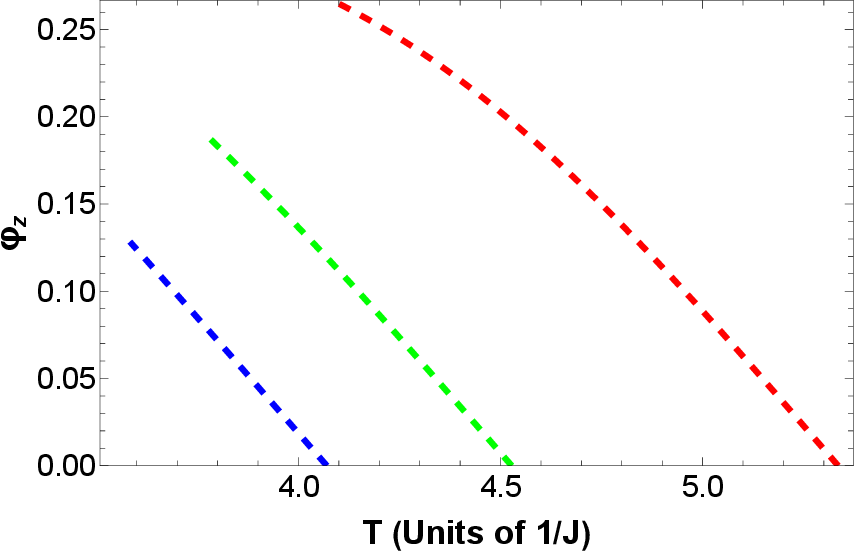}\label{fig:phiz_1}
\end{subfigure}
\hspace{.2cm}
\begin{subfigure}[b]{0.4\textwidth}
    \centering\caption{}\includegraphics[width=\linewidth]{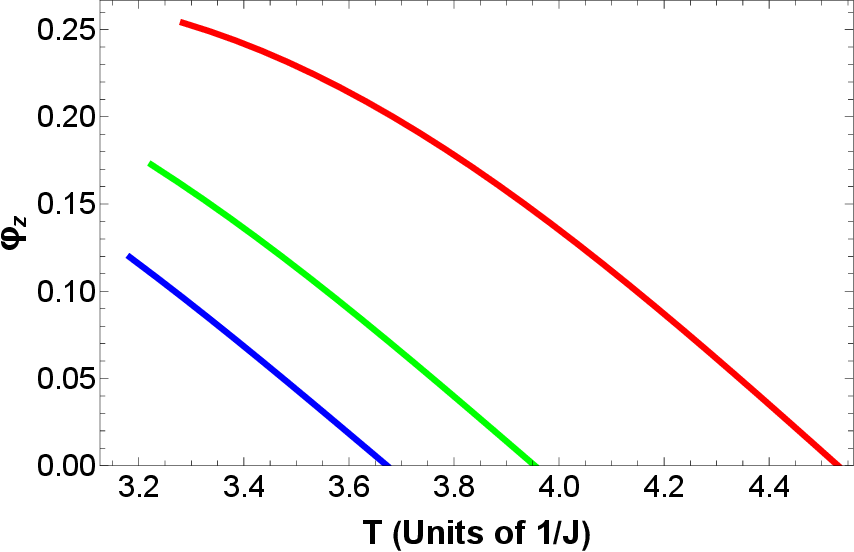}\label{fig:phiz_2}
\end{subfigure}
\caption{$\phi_z(\tau_1+\tau_2)$ corresponding to pulse-sequences I (a) and II (b), as a function of total duration, for parameters $\Omega_0/J=2.5$ (red lines), $\Omega_0/J=4$ (green lines) and $\Omega_0/J=6$ (blue lines).}
\label{fig:phi_z}
\end{figure}

In Fig. \ref{fig:phi_z} we plot $\phi_z(\tau_1+\tau_2)$ versus the charging duration $T$, for pulse-sequences I (Fig. \ref{fig:phiz_1}) and II (Fig. \ref{fig:phiz_2}), for the three values $\Omega_0/J = 2.5, 4, 6$ previously used (red, green and blue lines, respectively). First note that each curve starts from the threshold above which the corresponding pulse-sequence is optimal, see Fig. \ref{fig:stored_energy} and the corresponding discussion. Second, note that in almost all the cases it is $\phi_z(\tau_1+\tau_2)>0$. This sign is consistent with Eqs. (\ref{phi_TE}) and the maximum principle requirement that the optimal control maximizes the control Hamiltonian. Specifically, for pulse-sequence I, at $t=\tau_1+\tau_2$ the control switches from $0$ to $\Omega_0$ and, since $\phi_x(\tau_1+\tau_2)=0$ and $\phi_z(\tau_1+\tau_2)>0$, we see from Eq. (\ref{phiy}) that $\dot{\phi}_y(\tau_1+\tau_2)<0$. But $\phi_y(\tau_1+\tau_2)=0$, thus immediately after the switching it is $\phi_y<0$, leading through Eq. (\ref{phix}) to $\phi_x>0$, which is consistent with the choice $\Omega(t)=\Omega_0$ in the interval $\tau_1+\tau_2<t\leq T$ for maximizing the control Hamiltonian (\ref{H_c}).
Analogously, for pulse-sequence II, at $t=\tau_1+\tau_2$ the control switches from $0$ to $-\Omega_0$ resulting in $\phi_y>0$ and thus to $\phi_x<0$, consistent with the choice $\Omega(t)=-\Omega_0$ in the interval $\tau_1+\tau_2<t\leq T$ for maximizing the control Hamiltonian (\ref{H_c}). Finally note that $\phi_z(\tau_1+\tau_2)=0$ at the full charging duration for each case, as theoretically explained at the end of Sec. \ref{sec:full}.

\section{Conclusion}
\label{sec:conclusion}

We studied the problem of maximizing the stored energy for a given charging duration in a quantum battery consisted of two spins-$1/2$ with Ising interaction, initially in the spin-down state, using bounded transverse field control. We defined an equivalent optimal control problem on an effective two-level system and showed that, despite that a single bang pulse can quickly obtain considerable charging levels for relatively large upper control bound, higher levels of stored energy including complete charging are achieved by a bang-singular-bang pulse-sequence, with an Off middle singular pulse. We studied the problem for the case where the control is restricted to be non-negative as well as for the case where it can also take negative values in a symmetric domain, and derived transcendental equations from which the durations of the individual pulses in the bang-singular-bang pulse-sequence can be found. For the case of full charging we curiously found for the bang-singular-bang solution that the three ``switching" functions for the equivalent qubit problem become zero while the adjoint ket does not, as required by optimal control theory. As a future work, we plan to investigate the fast charging of the spin-pair quantum battery when an arbitrary phase is allowed in the controls. In this case the optimal control problem changes considerably. For example, for a piecewise constant phase with three arbitrary values the evolution cannot be mapped to that of the same two-level system, thus a separate treatment becomes imperative. Since more control functions become available, it is in principle possible to obtain solutions which achieve faster charging. We also plan to study analytically the fast charging of an Ising spin-chain with few spins, for example three or four, although a numerical optimization approach may eventually become inevitable due to the complexity of the problem.

%Additionally, the abnormal optimal solution contains a singular arc. We attributed this phenomenon to the requirement of the qubit equivalent problem on the global phase and not just the position of the Bloch vector on the sphere, which is usually the case. We would like to extend this work to spin chains with three or more qubits.

\appendix

\section{Short introduction to optimal control theory}

\label{sec:appendixA}

%Optimal control theory \cite{Pontryagin} was developed during the cold war to answer questions related to the space race, for example what is the minimum-fuel trajectory to the moon. Through the years it has found a broad range of applications in a variety of systems including aerospace, electrical, biomedical and nuclear, but also has been successfully applied in quantum systems.  In the quantum context, usually the goal is to find the electromagnetic field which can drive a quantum system from some initial state to a desired target state, in the minimum possible time or with maximum fidelity, reducing thus the unwanted effect of noise.

Consider the autonomous (not explicitly time-dependent) control system
\begin{equation}
\label{autonomous}
\dot{\mathbf{x}}=\mathbf{f}(\mathbf{x},\mathbf{u}),
\end{equation}
where $\mathbf{x}, \mathbf{u}$ are the real vectors of state and control variables, respectively. For quantum systems, the state variables are usually elements of the wavevector or the density matrix, while the control variables are the control fields, f.e. $\Omega(t)$ in the current example. Usually the goal is to find the controls in the interval $[0 \; T]$ which minimize a cost function
\begin{equation}
\label{cost}
J=\Phi(\mathbf{x}(T))+\int_0^T L(\mathbf{x},\mathbf{u})dt,
\end{equation}
where $\Phi, L$ are the terminal and running costs, respectively. The initial state $\mathbf{x}(0)$ is usually fixed, the terminal time $t=T$ may be free and determined from the optimization, while the terminal state $\mathbf{x}(T)$ may need to satisfy some terminal boundary conditions, including the case where it is also fully fixed.
For example, when the terminal state is fixed, the choice $\Phi=0, L=1$ minimizes the transfer time $T$ from the initial state $\mathbf{x}(0)$ to the target $\mathbf{x}(T)$, while the choice $\Phi=0, L=|\mathbf{u}|^2/2$ minimizes the energy of the control pulses for the same transfer. For fixed duration $T$ and target quantum state $|\psi_f\rangle$, the choice $\Phi=1-|\langle\psi(T)|\psi_f\rangle|^2, L=0$ maximizes the fidelity of the final quantum state $|\psi(T)\rangle$.

Pontryagin's Maximum Principle \cite{Pontryagin62} provides first order necessary conditions for optimality, which are obtained roughly speaking by applying calculus of variations on the cost augmented with the dynamical constraints (\ref{autonomous}) using appropriate Lagrange multipliers. According to this theorem,
the optimal control is such that it minimizes the \emph{control} Hamiltonian
\begin{eqnarray}
\label{Hc1}
H_c&=&L(\mathbf{x},\mathbf{u})+\boldsymbol{\lambda}^T\cdot\dot{\mathbf{x}} \nonumber \\
   &=&L(\mathbf{x},\mathbf{u})+\boldsymbol{\lambda}^T\cdot\mathbf{f}(\mathbf{x},\mathbf{u}),
\end{eqnarray}
where $\boldsymbol{\lambda}$ is the vector of adjoint variables (costates), which are essentially time-dependent Lagrange multipliers associated with the dynamical constraints (\ref{autonomous}), satisfying the adjoint equations
\begin{equation}
\label{adjoint}
\dot{\boldsymbol{\lambda}}^T = -\frac{\partial H_c}{\partial\mathbf{x}}.
\end{equation}
Observe that, since Eq.(\ref{autonomous}) can also be expressed as $\dot{\mathbf{x}}^T=\partial H_c/\partial\boldsymbol{\lambda}$, Eqs. (\ref{autonomous}) and (\ref{adjoint}) form a pair of Hamilton's equations for the control Hamiltonian (\ref{Hc1}). If the dimension of the state vector is $n$, then $2n$ integration constants are needed for the solution of this system. The fixed initial state $\mathbf{x}(0)$ provides $n$ integration constants, while the terminal costate conditions provide the other $n$ such constants
\begin{equation}
\label{T_adjoint}
\boldsymbol{\lambda}^T(T) = \frac{\partial\Phi}{\partial\mathbf{x}(T)},
\end{equation}
where note that Eq. (\ref{T_adjoint}) is valid in the absence of terminal state conditions, as is the case for the problem under study. 

For this problem, there is no running cost, $L=0$, while $\Phi$ is given in Eq. (\ref{A_Stored}). The (real) state variables are the real and imaginary parts of $A, B$ in Eq. (\ref{two_level}), leading to the control Hamiltonian (\ref{H_c}), while the corresponding costates satisfy Hamilton's Eqs. (\ref{costates}). If we properly group them to form the adjoint ket $\ket{\lambda(t)}=[\lambda_A(t), \lambda_B(t)]^T$, then it can be easily shown that it satisfies Eq. (\ref{lambda}). The terminal costate conditions (\ref{terminal_costates}) are obtained by direct application of Eq. (\ref{T_adjoint}) using the $\Phi$ given in Eq. (\ref{A_Stored}), and are compactly expressed in Eq. (\ref{final_lambda}). We close by pointing out that here we actually maximize $\Phi$, and for this reason we also maximize the control Hamiltonian.

\section{Parameters $u_I, u_z, x_I, x_z, y_I, y_z$ as functions of $\tau_s, \tau_d, T$}

\label{sec:appendixB}

Pulse-sequence I:

\begin{subequations}
\label{u_pos}
    \begin{eqnarray}
    u_I &=& \cos \left[\frac{J  \left( T-\tau_s \right)}{2}\right] \cos \frac{\omega \tau_s}{2} \nonumber \\
    &-& n_z \sin \left[\frac{J  \left( T-\tau_s \right)}{2}\right] \sin \frac{\omega \tau_s}{2},  \\
    u_z &=& - n_x^2 \sin \left[\frac{J  \left( T-\tau_s \right)}{2}\right] \cos \frac{\omega \tau_d}{2} \nonumber \\
    &-& n_z \cos \left[\frac{J  \left( T-\tau_s \right)}{2}\right] \sin \frac{\omega \tau_s}{2} \nonumber \\
    &-& n_z^2 \sin \left[\frac{J  \left( T-\tau_s \right)}{2}\right] \cos \frac{\omega \tau_s}{2}.
    \end{eqnarray}
\end{subequations}

\begin{subequations}
\label{xs_ys_equal}
    \begin{eqnarray}
    x_I &=& - n_x \sin \frac{\omega \tau_s}{2} \cos \left[\frac{J  \left(T-\tau_s\right)}{2} \right], \\
    x_z &=& n_x \sin \frac{\omega \tau_d}{2} \sin \left[\frac{J  \left(T-\tau_s\right)}{2} \right] \nonumber \\
    &-& n_x n_z \cos \frac{\omega \tau_d}{2} \cos \left[\frac{J  \left(T-\tau_s\right)}{2} \right] \nonumber \\
    &+& n_x n_z \cos \frac{\omega \tau_s}{2} \cos \left[\frac{J  \left(T-\tau_s\right)}{2} \right], \\
    y_I &=& - n_x \sin \frac{\omega \tau_s}{2} \sin \left[\frac{J  \left(T-\tau_s\right)}{2} \right], \\
    y_z &=& - n_x \sin \frac{\omega \tau_d}{2} \cos \left[\frac{J  \left(T-\tau_s\right)}{2} \right] \nonumber \\
    &-& n_x n_z \cos \frac{\omega \tau_d}{2} \sin \left[\frac{J  \left(T-\tau_s\right)}{2} \right] \nonumber \\
    &+& n_x n_z \cos \frac{\omega \tau_s}{2} \sin \left[\frac{J  \left(T-\tau_s\right)}{2} \right].
    \end{eqnarray}
\end{subequations}

Pulse-sequence II:

\begin{subequations}
\label{u_sym}
    \begin{eqnarray}
    u_I &=& n_x^2 \cos \left[\frac{J  \left(T-\tau_s\right)}{2} \right] \cos \frac{\omega \tau_d}{2} \nonumber \\
    &-& n_z \sin \left[\frac{J  \left(T-\tau_s\right)}{2} \right] \sin \frac{\omega \tau_s}{2} \nonumber \\
    &+& n_z^2 \cos \left[\frac{J  \left(T-\tau_s\right)}{2} \right] \cos \frac{\omega \tau_s}{2}, \\
    u_z &=& - \sin \left[\frac{J  \left(T-\tau_s\right)}{2} \right] \cos \frac{\omega \tau_s}{2} \nonumber \\
    &-& n_z \cos \left[\frac{J  \left(T-\tau_s\right)}{2} \right] \sin \frac{\omega \tau_s}{2}.
    \end{eqnarray}
\end{subequations}

\begin{subequations}
\label{xs_ys_inverted}
    \begin{eqnarray}
    x_I &=& - n_x \sin \frac{\omega \tau_d}{2} \cos \left[\frac{J  \left(T-\tau_s\right)}{2} \right] \nonumber \\
    &-& n_x n_z \cos \frac{\omega \tau_d}{2} \sin \left[\frac{J  \left(T-\tau_s\right)}{2} \right] \nonumber \\
    &+& n_x n_z \cos \frac{\omega \tau_s}{2} \sin \left[\frac{J  \left(T-\tau_s\right)}{2} \right], \\
    x_z &=& n_x \sin \frac{\omega \tau_s}{2} \sin \left[\frac{J  \left(T-\tau_s\right)}{2} \right], \\
    y_I &=& - n_x \sin \frac{\omega \tau_d}{2} \sin \left[\frac{J  \left(T-\tau_s\right)}{2} \right] \nonumber \\
    &+& n_x n_z \cos \frac{\omega \tau_d}{2} \cos \left[\frac{J  \left(T-\tau_s\right)}{2} \right] \nonumber \\
    &-& n_x n_z \cos \frac{\omega \tau_s}{2} \cos \left[\frac{J  \left(T-\tau_s\right)}{2} \right], \\
    y_z &=& - n_x \sin \frac{\omega \tau_s}{2} \cos \left[\frac{J  \left(T-\tau_s\right)}{2} \right].
    \end{eqnarray}
\end{subequations}

\begin{acknowledgements}
The present work was financially supported by the ``Andreas Mentzelopoulos Foundation". The work of D.S. was funded by an Empirikion Foundation research grant. D.S. would like to acknowledge insightful discussions with Dominique Sugny.
\end{acknowledgements}

%\appendix*

%\section{If Needed}

\bibliographystyle{apsrev4-2}
\bibliography{main}

\vspace*{1. cm}

\end{document}